\documentclass[aps, amsmath, reprint, 10pt]{revtex4-1} 

\usepackage{pgfplots}     
\usepackage{anyfontsize}  
\pgfplotsset{compat=1.13}
\usepackage{graphicx}
\usepackage{hyperref}     
\graphicspath{{figures/}} 

\usepackage{color,soul}

\newcommand{\ls}[0]{\lambda_s}
\newcommand{\lsc}[0]{\lambda_s^c}
\newcommand{\ld}[0]{\lambda_d}
\newcommand{\ldi}[0]{\lambda_{d}}

\newcommand{\vnz}[0]{\textrm{NZPV}}

\begin{document}

\title{Concepts of Static vs. Dynamic Current Transfer Length in 2G HTS coated conductors with a Current Flow Diverter Architecture}


\author{Jean-Hughes Fournier-Lupien}
\affiliation{Polytechnique Montréal, Montréal (QC) H3C 3A7, Canada}
\affiliation{Karlsruhe Institute of Technology, Institute of Technical Physics, 76344 Eggenstein-Leopoldshafen, Germany}

\author{Frédéric Sirois}
\affiliation{Polytechnique Montréal, Montréal (QC), H3C 3A7, Canada}
\affiliation{Karlsruhe Institute of Technology, Institute of Technical Physics, 76344 Eggenstein-Leopoldshafen, Germany}

\author{Christian Lacroix}
\affiliation{Polytechnique Montréal, Montréal (QC), H3C 3A7, Canada}

\date{\today}

\begin{abstract}
This paper uses both experimental and numerical approaches to revisit the concept of current transfer length (CTL) in second-generation high-temperature superconductor coated conductors with a current flow diverter (CFD) architecture. The CFD architecture has been implemented on eight commercial coated conductors samples from THEVA. In order to measure the 2-D current distribution in the silver stabilizer layer of the samples, we first used a custom-made array of 120 voltage taps to measure the surface potential distribution. Then, the so-called ``static'' CTL ($\ls$) was extracted using a semi-analytical model that fitted well the experimental data. As defined in this paper, the static CTL on a 2-D domain is a generalization of the definition commonly used in literature. In addition, we used a 3-D finite element model to simulate the normal zone propagation in our CFD samples, in order to quantify their ``dynamic'' CTL ($\ld$), a new concept introduced in this paper and defined as the CTL observed during the propagation of a quenched region. The results show that, for a CFD architecture, $\ld$ is always larger than $\ls$, whereas $\ld = \ls$ when the interfacial resistance between the stabilizer and the superconductor layers is the same everywhere. We proved that the cause of these different behaviors is related to the shape of the normal zone, which is curved for the CFD architecture, and rectangular otherwise. Finally, we showed that the NZPV is proportional to $\ld$, not with $\ls$, which suggests that the dynamic CTL $\ld$ is the most general definition of the CTL and should always be used when current crowding and non-uniform heat generation occurs around a normal zone. 
\end{abstract}

\pacs{}
\maketitle

\section{Introduction}

Second-generation high-temperature superconductor coated conductors (2G HTS CCs) based on REBa$_2$Cu$_3$O$_{7-x}$ (RE = rare earth), sometimes simply called HTS tapes, are expected to be used in many applications where high current density, high field or high mechanical strength is required~\cite{Trociewitz2011,Senatore2014,Maeda2019,Uglietti2019}.
However, whatever application is targeted, HTS tapes are prone to develop hot spots, and these hot spots can easily destroy the tape (and even the device that comprises the tape) if no action is taken rapidly. The most common protection action is to cut the current when a hot spot starts to propagate along the tape. A fast propagation of the hot spot, also called normal zone propagation velocity (NZPV), is generally desirable, since it allows detecting rapidly the hot spot using the a voltage rise across the terminals of the tape, often winded as a coil.

The NZPV of a given tape architecture can be measured experimentally, but it can also be deduced from another physical quantity: the current transfer length (CTL)~\cite{Levin2010a,Lacroix2014a}. In~\cite{Levin2007b}, Levin showed that there is a linear relationship between the NZPV and the CTL for classical HTS tape architectures (also called ``uniform architecture'' in this work), i.e. tapes with a 2-D layered structure and sufficiently high interfacial resistance between the REBCO layer and the silver stabilizer. Provided that the CTL can be easily measured or even determined analytically for such tape architectures, it suggests that the CTL is a good indicator of tape robustness against hot spots: the longer it is, the larger is the NZPV. The CTL is thus an important characteristic of superconducting tapes and wires, and it has been intensively studied over the years for multi-filamentary low temperature superconductors~\cite{Ekin1978,Dhalle1999,Polak1997}, for MgB$_2$ wires~\cite{Stenvall2007, holubek2007, vinod2010} and of course for 2G HTS CCs~\cite{Polak2006,Duckworth2009, Bagrets2016}.

Despite the existence of many investigations related to uniform superconducting wire architectures~\cite{Fang1996,Polak1997,Duckworth2001,Usoskin2002,Levin2007b}, in which the definition of the CTL was based on the theory developed for low temperature superconductors~\cite{yu1982,Wilson1983}, the relationship between the NZPV and the CTL has never been established for HTS tape architectures comprising a current flow diverter (CFD)~\cite{Lacroix2014b,Lacroix2014a,Lacroix2017}, an architecture devised to substantially speed up the NZPV and make quench protection easier.

In this paper, we address explicitly this question using a mixed experimental and numerical simulation methodology, which led us to define two types of CTL, namely a ``static CTL'', which is compatible with the classical definition, and a ``dynamic CTL'', which is shown to be more general than the classical static definition.


\section{Experimental approach}
\label{section:experiment}
\subsection{Preparation of samples} 
The HTS tapes used in this work were 2G HTS CCs fabricated by THEVA~\cite{THEVA}. The fabrication process was as follows. A 100~$\mu$m thick Hastelloy (C-276$^{\textrm{TM}}$) tape was used as the substrate, after being mechanically grinded and electro-chemically polished to smooth its surface. Then, a 3~$\mu$m thick film of MgO was deposited by evaporation (e-gun) and textured at the same time. The texturing was obtained with an inclined substrate deposition technique (ISD~\cite{Bauer1999,Bauer2000,Metzger2001,Zilbauer2006}) in an oxygen atmosphere. The film was then capped with an additional 200~nm thick layer of MgO deposited at straight incidence. After that, a 3~$\mu$m thick layer of GdBaCuO was grown on top of the MgO layer using thermally reactive co-evaporation~\cite{Kinder1997}. Finally, a 1~$\mu$m thick silver layer was evaporated all around the tape. All samples used in this work were 5 cm long.

\begin{figure}[t]
	\centering
	\includegraphics[width=\columnwidth, trim={0cm -1cm 0cm 0cm},clip ]{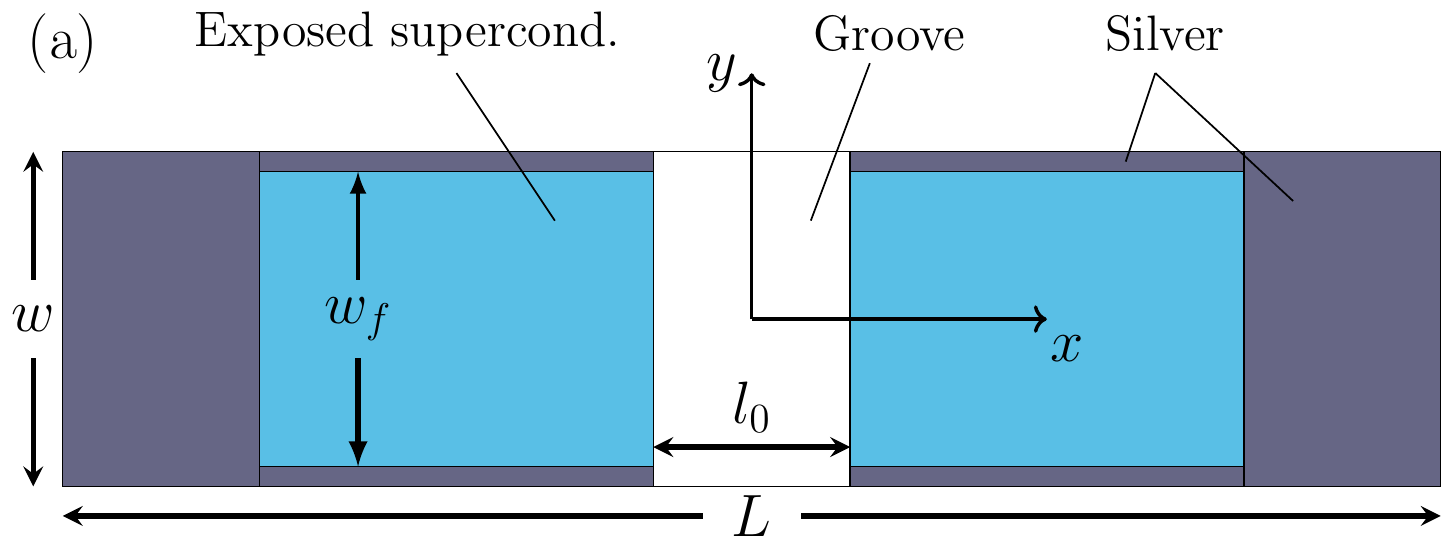}
	\includegraphics[width=\columnwidth, trim={0cm 0cm 0cm 0cm}, clip ]{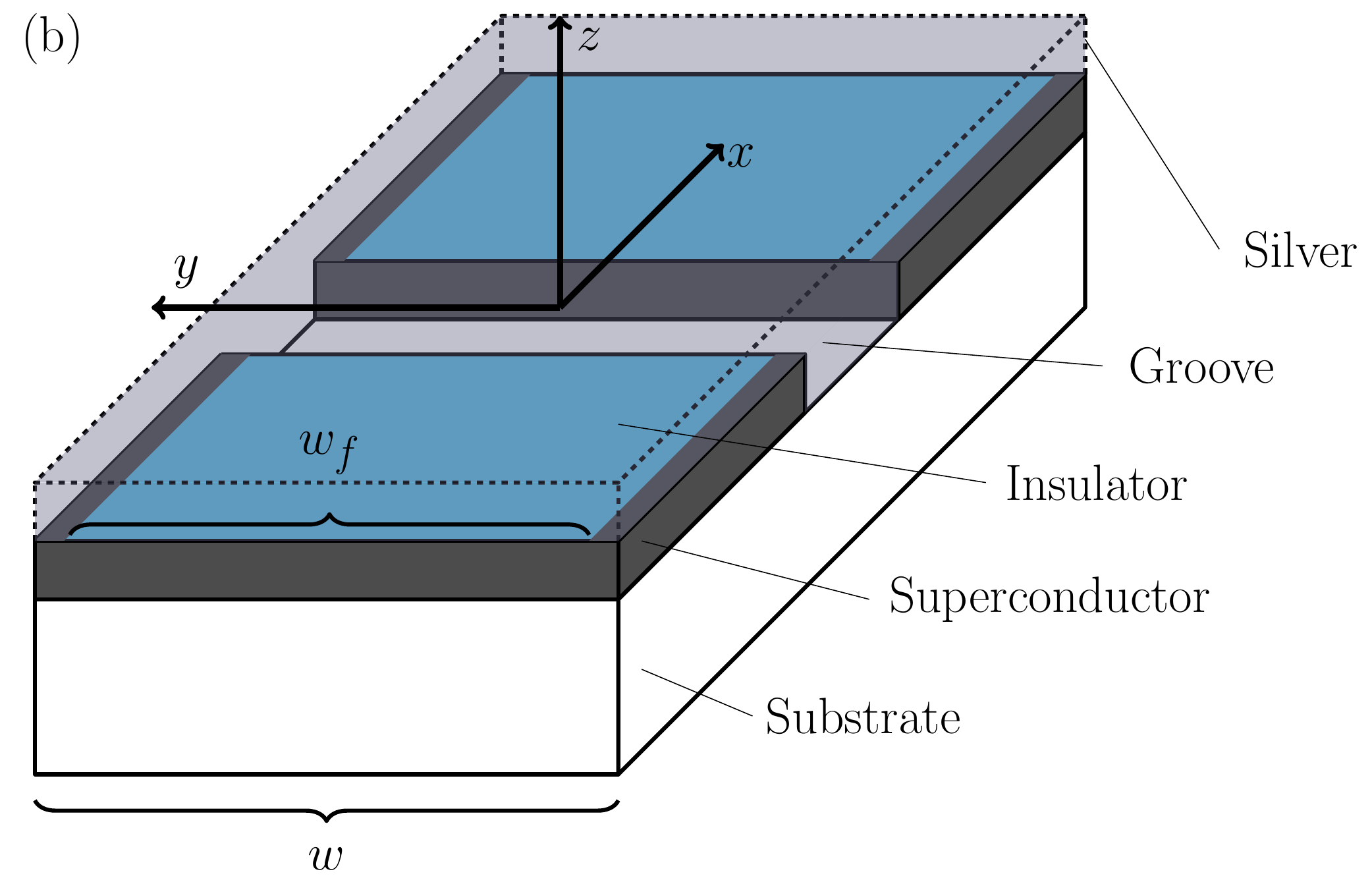}
	\caption{(a) Top view of a sample after the chemical etching operations required to fabricate the CFD (thin amorphous superconducting layer, in cyan) and the groove in the superconductor. (b) 3-D view of the tape architecture, showing that the silver stabilizer also fills the groove.}
	\label{fig:fig1}
\end{figure}

These samples had to be further modified to obtain a ``current flow diverter'' (or CFD) architecture~\cite{Lacroix2014a,Lacroix2014b}. This was done by etching the silver in the center of the tape while keeping the silver intact on its edges, as shown in figure~\ref{fig:fig1}(a). When etching the silver, the superconductor becomes exposed to CO$_2$ and water vapor in the air, which creates a very thin amorphous layer (in cyan) that is electrically resistive~\cite{Russek1994,Ekin2006,Lacroix2014a}. The width of the etched silver part, noted $w_f$, was varied for each sample. In total, eight samples with different $w_f$ values varying from 0.0 to 12.0 mm were produced.
The silver layer was kept intact (no etching) at the current contacts.

In order to be able to measure the CTL easily, we modified the samples to force the current transfer between the silver and the superconductor layer. The first step was to etch completely the silver on the substrate side and on the edges of the tape. The second step was to create a groove of width $l_0$ in the silver and the superconductor layers across the whole width $w$ of the tape, as illustrated in figure~\ref{fig:fig1}(a). The silver was etched with a solution of ammonium hydroxide and peroxide diluted in water in a volume ratio of 1:1:4 respectively, then the groove was deepened by etching the superconductor layer with phosphoric acid. The targeted groove width $l_0$ was 10~mm for all samples. Finally, a new silver layer between 1 and 2~$\mu$m of thickness was sputtered on top of each sample. The final architecture of the samples prior to the electrical characterization is shown in figure~\ref{fig:fig1}(b).

After all these fabrication steps, we obtained a set of eight HTS tapes with a CFD architecture and a groove at their middle length. The property of a CFD tape is to exhibit a high electrical resistance at the interface between the superconductor and the silver layers (cyan surface in figure~\ref{fig:fig1}), noted $R_f$ $(\mu\Omega.$cm$^2)$, and a low electrical resistance at the interface between the superconductor and the silver layers at the edges of the tape, called $R_0$ $(\mu\Omega.$cm$^2)$. The values of the interfacial resistances $R_f$ and $R_0$ were measured on two special samples. To measure $R_f$, a sample with the whole superconductor layer exposed to air was used, whereas to measure $R_0$, we simply used a virgin THEVA sample. The values found were respectively $R_f=10^{3}~\mu\Omega.$cm$^2$ and $R_0=10^{-1}~\mu\Omega.$cm$^2$. These measurements were done with the technique described in~\cite{Fournier-Lupien2020a, Lacroix2013a, Polak2006}.

\subsection{CTL measurement setup}
\begin{figure}[bt]
	\centering
	\includegraphics[ width=\columnwidth, trim={0cm 0cm 0cm 0cm},clip ]{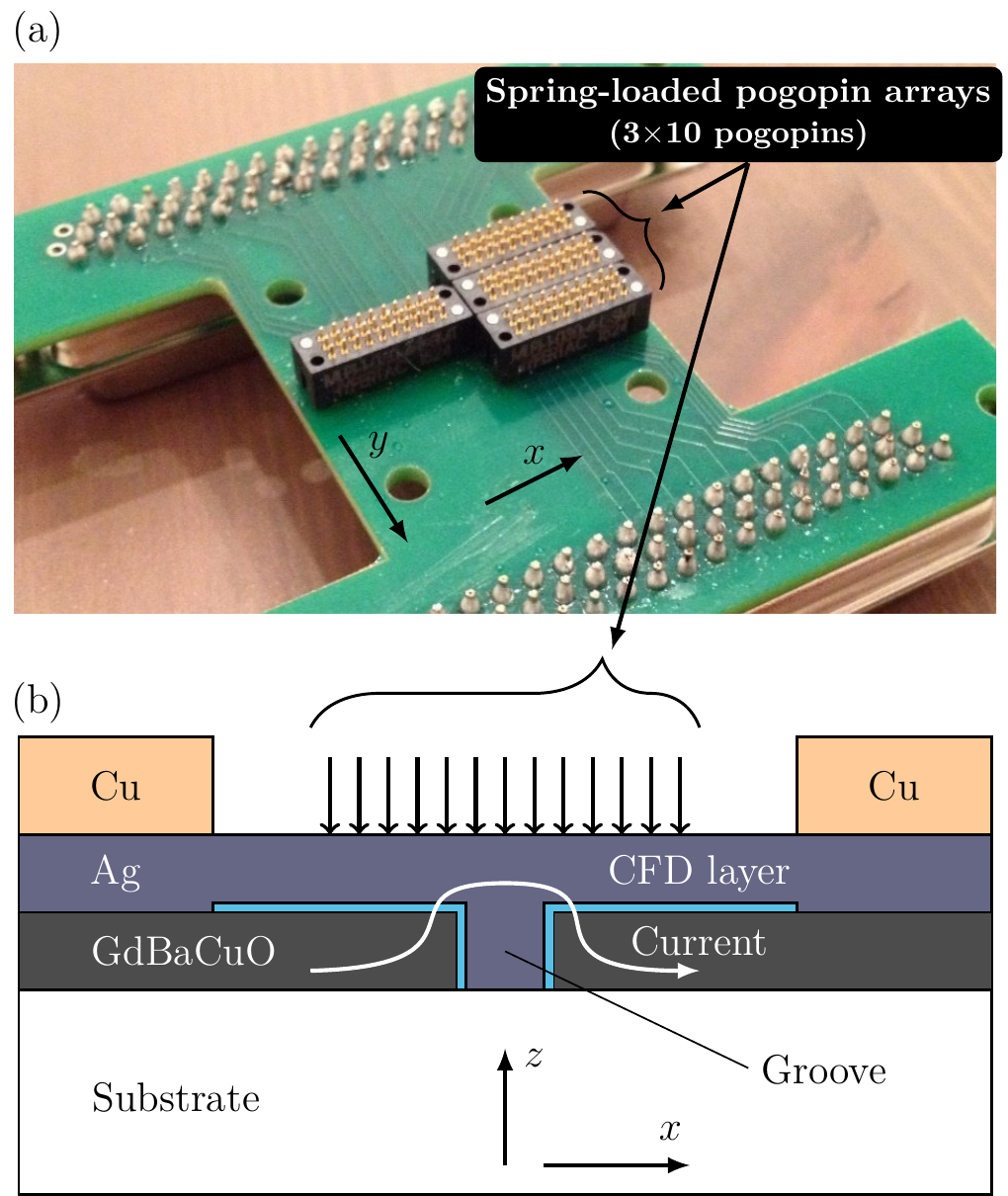}
	\caption{ (a) Picture of the four arrays of $3\times10$ pogo pins soldered on a PCB. (b) Schematic view of the sample cross-section, showing the current path around the groove (not to scale). The black arrows above the groove represent the pogo pins.}
	\label{fig:fig2}
\end{figure}

The CTL can be deduced from surface potential measurements performed on a tape, just above the region where current transfers from superconductor to silver or vice-versa. We thus designed and built a homemade sample holder able to map the surface potential in two directions: along the length of the tape and across its width. This 2-D in-plane mapping was required because of the 3-D path followed by the current in the transfer region when considering a CFD architecture (explanations provided later). In this work, four arrays of $3\times10$ spring-loaded gold-plated pogo pins~\footnote{Hypertac MLPI-30-4-L-SMT} with a one millimeter spacing were used, for a total of 120 pogo pins. The four arrays of pogo pins were soldered on a 6-layer PCB, as shown in figure~\ref{fig:fig2}(a). For performing a measurement, the pogo pins must be placed above the groove and pressed against the sample. 

During the experiments, the sample holder was immersed in liquid nitrogen. The current was injected in the sample using a Keithley 6221 current source via two copper blocks pressed against the sample at both extremities (where the silver has not been etched), see figure~\ref{fig:fig2}(b). The injected current was kept low (100~mA) in order to limit the generation of heat. 

Using a Keithley 2182A nanovoltmeter, the voltage was measured between a selected pogo pin and a reference pogo pin, located far away from the groove. The voltage of the reference pogo pin was considered to be $V=0$. The voltage was measured using the delta method~\cite{Keithley2004} in order to cancel any thermoelectric voltage that could be present.

\section{Experimental results} \label{Exp_CTL}

\subsection{Measurement of surface electric potential}

\begin{figure}[tb] 
	\centering
	\includegraphics[width=\columnwidth, trim={ 0cm 0cm 0cm 0cm},clip]{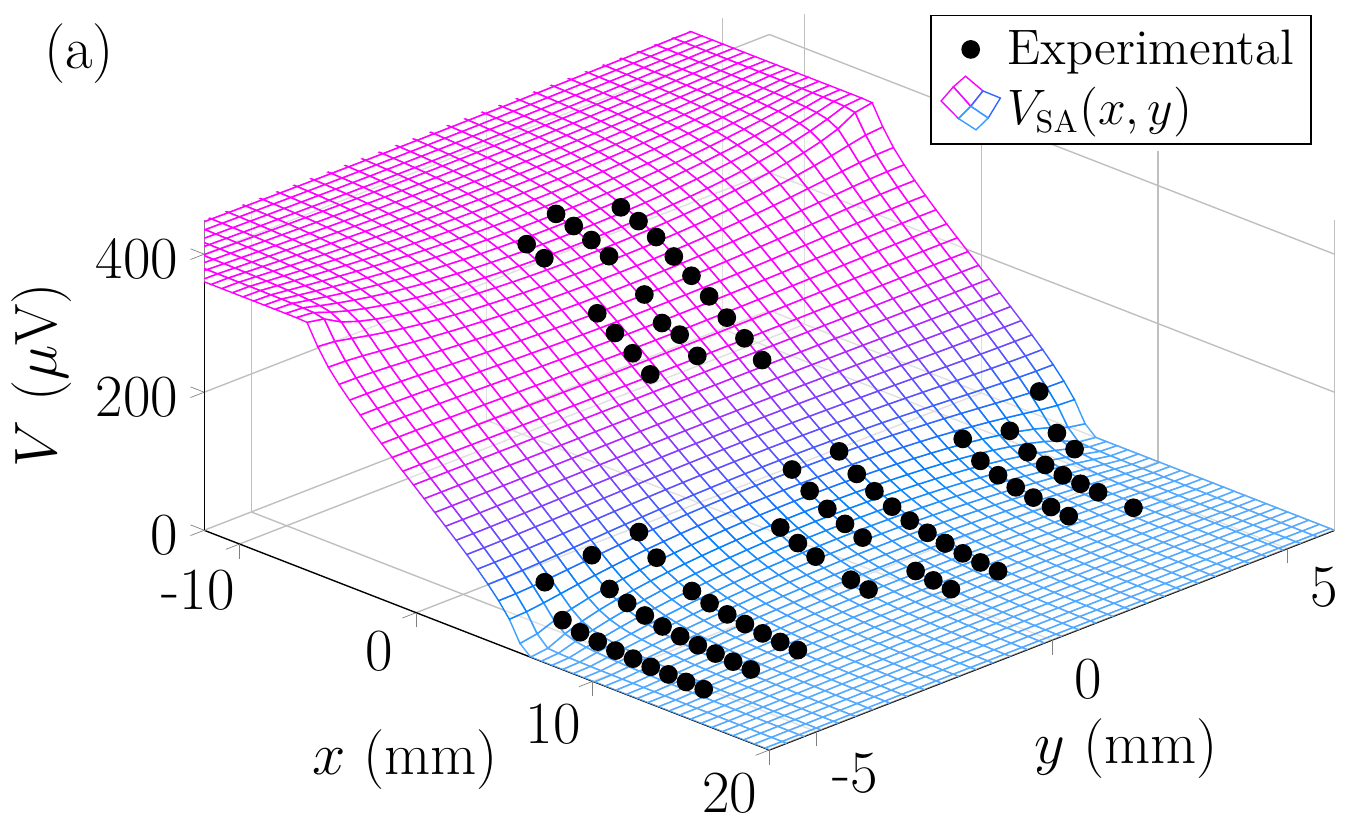}	
	\includegraphics[width=\columnwidth, trim={ 0cm 0cm 0cm 0cm},clip]{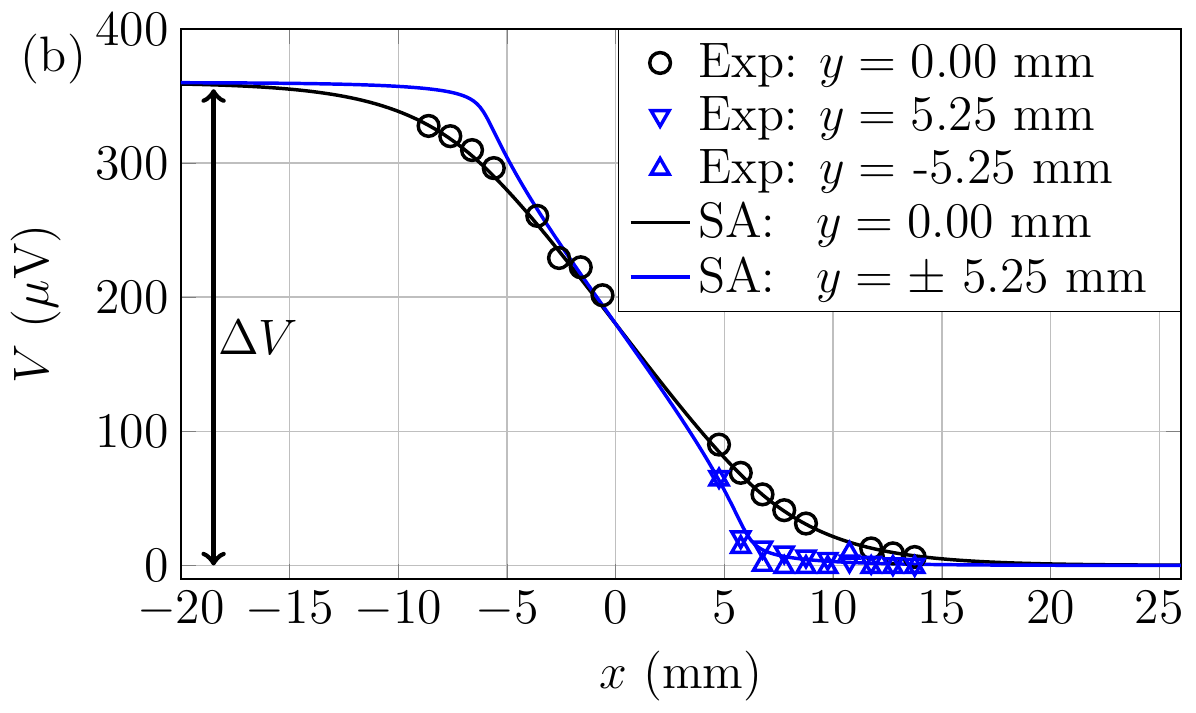}	
	\includegraphics[width=\columnwidth, trim={ 0cm 0cm 0cm 0cm},clip]{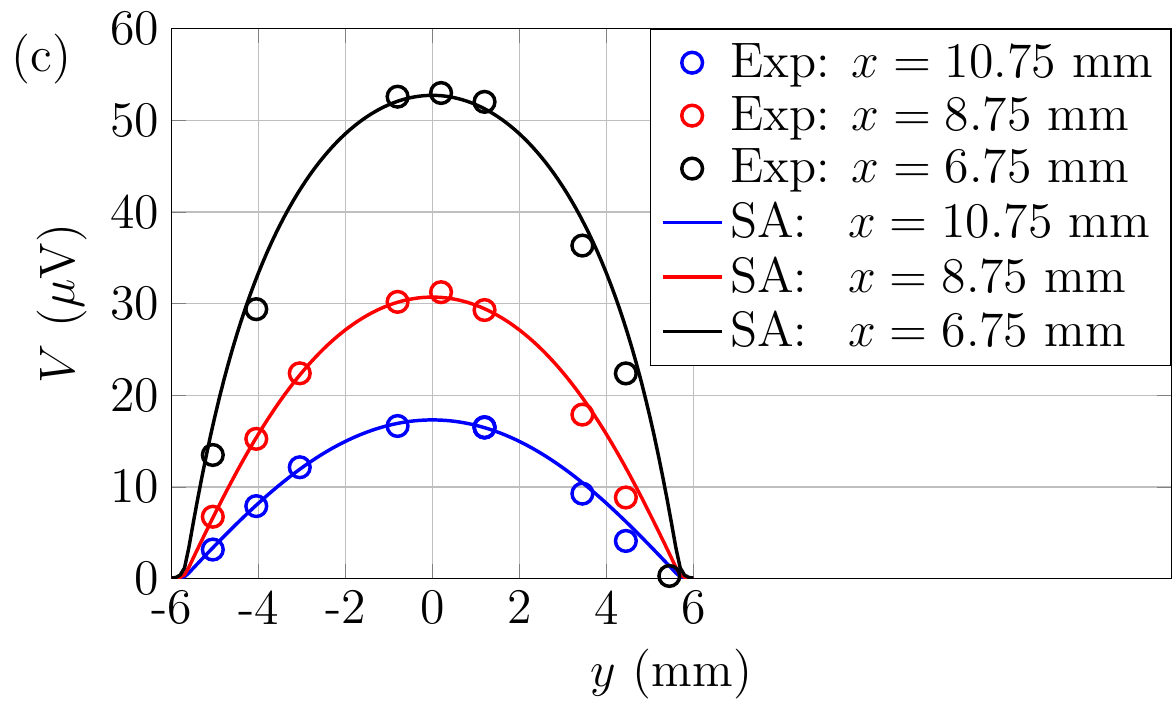}	
	\caption{
Typical surface potential data $V(x,y)$ measured experimentally (symbols) on a CFD sample with a groove (sample 5 in Table~\ref{tab:optimi}). Straight lines correspond to the potentials calculated with the semi-analytical (SA) model presented in next section. (a) All available experimental data (solid black circles) plotted on a 3-D calculated surface (wireframe surface). (b) Slices of $V(x,y)$ along the $x$ direction. (c) Slices of $V(x,y)$ along the $y$ direction.
}
	\label{fig:fig3}
\end{figure}

The experimental surface potential distribution $V(x,y)$ was measured on the silver layer for all eight samples with a groove. A typical surface potential measurement with the 120 pogo pins is presented in figure~\ref{fig:fig3}(a) (solid black circles). We observe that some data points are missing: this is due to some faulty connections between the pogo pins and the cables connected to the PCB. Figure~\ref{fig:fig3}(b) and figure~\ref{fig:fig3}(c) show the same data displayed along $x$ and $y$ (open symbols). In these figures, the groove is centered at $x=0$.

\subsection{Semi-analytical model of the electrical potential}

The experimental data were fitted with a semi-analytical model (SA) in order to get a clearer view of the surface potential profile at all ($x$,$y$) points. Knowing the electrical resistivity of silver, one can deduce the local current density everywhere in the silver layer, as well as the total net current flowing in silver, and if necessary, the total net current in the superconductor as well, knowing that no current flows in the Hastelloy substrate due to the electrical insulation properties of the MgO buffer layer. The current in silver is required to obtain the CTL, as described in the next sections.

The detailed development of the SA model can be found in~\cite{Fournier-Lupien2020a}. Only the main points are summarized below. The model assumes that both the silver and the superconductor layers are thin enough to neglect variations of the potential along their thickness $z$, which leads to a description of the problem in terms of the planar Helmholtz equation~\cite{Levin2007b}, i.e.
\begin{equation} \label{eq:prob}
	\nabla^2 V(x,y) - \kappa^2(y) \left( V(x,y) - V_s(x,y) \right) = 0,
\end{equation}
where $V(x,y)$ is the surface potential of the silver layer, $V_s(x,y)$ is the potential in the superconducting layer,
$\kappa (y)= \sqrt{\frac{\rho_{\textrm{Ag}}}{R(y) d_{\textrm{Ag}} }}$, where $\rho_{Ag}$ is the conductivity of silver and $R(y)$ is the local interfacial resistance, equal to either $R_f$ or $R_0$. According to figure~\ref{fig:fig1}, we can write 
\begin{equation} \label{eq:R}
R(y)=
\begin{cases} 
	R_{f}, &~\textrm{if}~ -w_f/2 < y < w_f/2 \,,   \\
	R_{0}, &~\textrm{if}~~\, \phantom{-}w_f/2 < |y| < w/2 \,.   \\
\end{cases}
\end{equation}

The developments in~\cite{Levin2007b} leads to an exact solution for $V(x,y)$, which we shall note $V_{\textrm{SA}}(x,y)$, expressed as a double infinite summation, i.e.
\begin{equation} \label{eq:solution}
V_{\textrm{SA}}(x, y) = \sum_{i = 1}^{\infty}{\sum_{j = 1}^{\infty}{c_{ij} \psi_{ij}(x, y)}} \,,
\end{equation}
where $\psi_{ij}(x, y)$ are analytical functions (eigenfunctions of a modified Laplace operator) and the coefficient $c_{ij}$ are determined numerically, which explains why the model is qualified as ``semi-analytical''.

Once $V_{\textrm{SA}}(x, y)$ is known, it is possible to find the net current $I(x)$ flowing in the silver layer as follows:
\begin{equation} \label{eq:I}
	I(x) = d_{Ag} \int_{-w/2}^{w/2}{J_x(x, y)  \,\textrm{d}y} \,,
\end{equation}
where $d_{Ag}$ is the thickness of the silver layer, and $J_x(x,y)$ is the local current density flowing along the length of the tape, given by
\begin{equation} \label{eq:J}
J_x(x,y) =  -\rho_{\textrm{Ag}}^{-1} \, \frac{\partial V_{\textrm{SA}}(x, y)}{\partial x }.
\end{equation}
Therefore, if the fitting procedure of the SA model with the experimental data is done properly, we have all we need to determine the experimental CTL in our samples.

\subsection{Fitting procedure}
Seven constant input parameters are required for evaluating $V_{\textrm{SA}}(x,y)$~\cite{Fournier-Lupien2020a}. These are $R_f$, $R_0$, $w$, $d_{\textrm{Ag}}$, $w_f$,  $l_0$, $\Delta V$. All of them were defined above, except $\Delta V$, which corresponds to the total voltage drop resulting from the presence of the groove, see figure~\ref{fig:fig3}(b). The first four parameters were measured experimentally. The values used in the SA model were $R_f= 10^{3}~\mu\Omega.$cm$^2$, $R_0=10^{-1}~\mu\Omega.$cm$^2$ and $w=12$~mm. The value of $d_{Ag}$ was $2.0~\mu$m for samples 2, 4 and 5, and $1.1~\mu$m for samples 1, 3, 6, 7 and 8.

The last three parameters, i.e.~$w_f$,  $l_0$ and $\Delta V$, were taken as degrees of freedom for minimizing the least-mean-square error between the SA model and the experimental data points for each sample. The results of this optimization is summarized in table~\ref{tab:optimi}. In fact, $w_f$ (width of CFD layer) was also measured experimentally, but since it was hard to be accurate in the fabrication and in the measurement, it was still left free to vary during the optimization, even if the optimizer did not change its value significantly. We thus concluded that the main degrees of freedom for this optimization were $l_0$ and $\Delta V$.

Regarding $l_0$ (width of the groove), the different values obtained account for the inaccuracies induced by the two-step etching process of the silver and superconductor layer. As for $\Delta V$, it represents a real fitting parameter, since we had no direct measurement of it. An example of the best fit of $V_{\textrm{SA}}(x,y)$ for sample 5 is presented in figure~\ref{fig:fig3}. The fit of the SA model with the experimental data worked well on all samples.

\begin{table}[tb] 
		\caption{Values of the best fitting parameters used with the SA model to reproduce the experimental data.
}
		\centering
		\begin{tabular}{cc|ccc}
				\hline \vspace{-3ex}   \\
				& \multicolumn{1}{c}{Measured } & \multicolumn{3}{c}{Optimized values} \\
			   \hline \vspace{-3ex}    \\
		       Sample & \phantom{----} $w_f$ \phantom{----} & $w_f$  & $l_0$ & $\Delta V$  \\
			        \# & (mm) & (mm)  & (mm)  & ($\mu$V)   \\
		       \hline \vspace{-3.5ex}  \\  
		       1 & 0.00 & 0.00  & 10.7 & 525  \\ 
		       2 & 9.48 & 9.10  & 11.5 & 360  \\ 
		       3 & 9.50 & 9.50  & 10.5 & 390  \\ 
		       4 & 10.80 & 10.93 & 10.0 & 380 \\ 
		       5 & 11.30 & 11.28 & 12.0 & 360 \\ 
		       6 & 11.50 & 11.40 & 8.6  & 350 \\ 
		       7 & 11.64 & 11.64 & 11.1 & 320 \\ 
		       8 & 12.00 & 12.00 & 10.5 & 1525\\ 
		       \hline \\   
		\end{tabular}
		\label{tab:optimi}
\end{table}

\subsection{Determination of the experimental CTL}
\label{CTL_exp} 

Once the electric potential $V_{\textrm{SA}}(x,y)$ is known everywhere in the silver layer, one can use (\ref{eq:I})~and~(\ref{eq:J}) to determine $I(x)$, i.e.~the total current carried by the silver layer at any point $x$ along the length of the tape. Note that $I(0)$ corresponds to the current in the middle of the groove, where there is no superconductor. Since silver is the only conducting path at this location, $I(0)$ also corresponds to the total current injected in the sample.
 
\begin{figure}[tb] 
	\centering	
	\includegraphics[width=\columnwidth, trim={-0.5cm 0cm 0cm 0cm},clip]{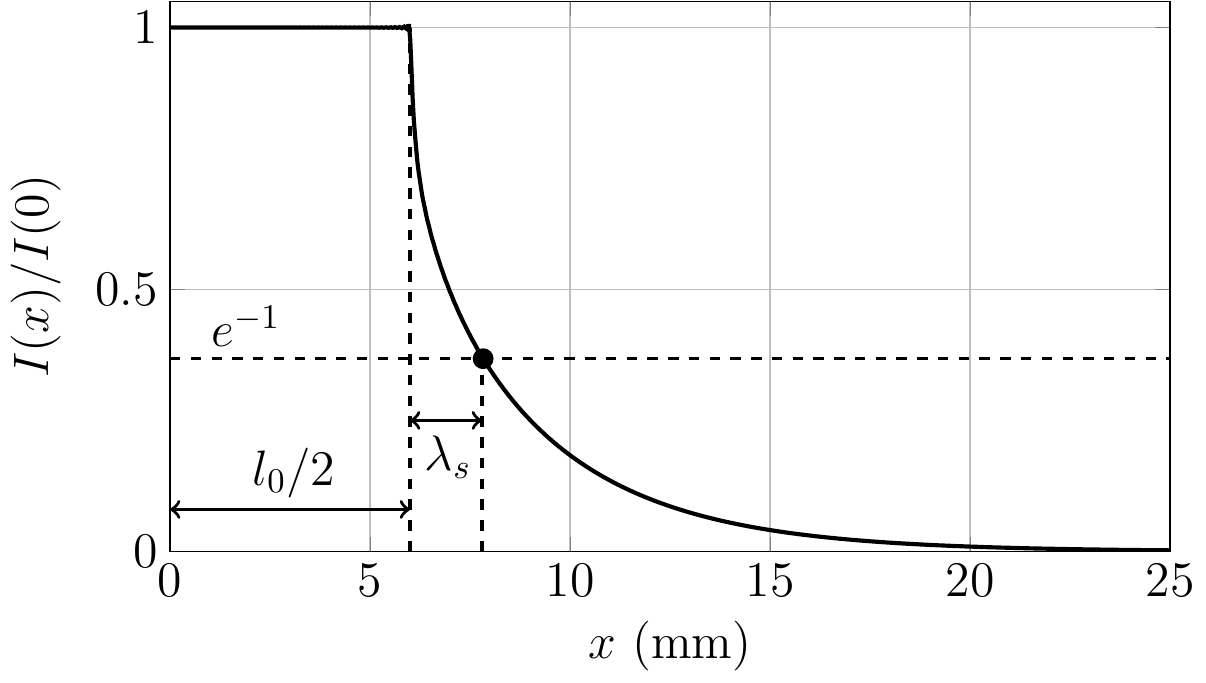}	
	\caption{
Fraction of injected current that flows in the silver layer at each point $x$ along the length of sample 5, as calculated with the SA model, after curve fitting with the experimental data. The current $I(x)$ was obtain with equations (\ref{eq:I}) and (\ref{eq:J}).
 }
	\label{fig:fig4}
\end{figure}

Figure~\ref{fig:fig4} shows the current obtained from equation~(\ref{eq:I}), considering the solution $V_{\textrm{SA}}(x,y)$ shown in figure~\ref{fig:fig3}
(data measured on sample 5). From its classical definition, the CTL corresponds to the distance $\ls$ from the edge of the groove where the current in the metallic shunt (here the silver layer) drops by a factor of $e^{-1}$, i.e. $I(\ls) / I(0) = e^{-1}$. In other words, $\ls$ is the characteristic length  required for the current to transfer from the stabilizer to the superconductor layer. A more detailed discussion about this definition can be found in~\cite{Fournier-Lupien2020a}. 

The CTL values determined with this definition for each of the eight samples produced in this work are given in table~\ref{tab:lambda}. Since the CTL depends on the thickness of the silver layer, and since the latter was not the same for all samples, we used the fitted values $w_f$, $l_0$ and $\Delta V$ for each sample and recalculated their CTL, assuming
a fixed thickness of $1~\mu$m for the silver layer, which gave a ``corrected'' value noted $\lsc$ in table~\ref{tab:lambda}. We use this value later in the paper for comparing our experiments with numerical simulations performed on tapes with 1-$\mu$m of silver stabilizer. After the correction, we observe that
$\lsc$ systematically increases as $f$ increases, where $f$ is the relative width of the CFD layer. This was the expected behavior.

Note that the ``$s$'' subscript in $\ls$ (or $\lsc$) stands for ``static CTL'', meaning that the current transfer happens at a fixed location over time, here at the edges of the groove. We show later that this CTL can be different from that observed at the extremities of an expanding normal zone (NZ), which we shall call a ``dynamic CTL'', noted $\ld$. Indeed, when a NZ (i.e.~hot spot) appears in a HTS tape carrying a current, the latter is forced to transfer in the silver layer to bypass the NZ.
If the current generates a sufficient amount of heat,
the NZ starts to expand at a specific velocity, called normal zone propagation velocity (NZPV), and the region over which the current transfer occurs follows the front of the NZ.

\begin{table}[tb] 
		\caption{
Summary of the various experimental and simulated values of the CTL and other quantities considered in this paper (NZPV stands for normal zone propagation velocity).
}
		\centering
		\begin{tabular}{ccc|cc|ccc}
				\hline \vspace{-3ex} \\
		        \multicolumn{3}{c}{ } &  \multicolumn{2}{c}{Experimental} & \multicolumn{2}{c}{FE1 model} \\
		       \hline
			   Sample & $f=w_f/w$ & $d_{\textrm{Ag}}$ & $\ls$ & $\lsc$ &  $\ldi$ & $\vnz$ \\
			   \# & no unit & ($\mu$m) & (mm) & (mm) & (mm) & (cm/s) \\
		       \hline \vspace{-3.5ex} \\  
		       1  &  0.00  & 1.1 & 0.07  &  0.06  &  0.08  &  4.8  \\
		       2  &  0.76  & 2.0 & 0.76  &  0.71  &  1.70  &  23.8 \\
		       3  &  0.79  & 1.1 & 0.87  &  0.87  &  2.00  &  27.1 \\
		       4  &  0.91  & 2.0 & 1.62  &  1.52  &  3.30  &  42.8 \\
		       5  &  0.94  & 2.0 & 1.83  &  1.72  &  3.69  &  47.8 \\ 
		       6  &  0.95  & 1.1 & 1.80  &  1.78  &  3.81  &  49.1 \\
		       7  &  0.97  & 1.1 & 1.97  &  1.95  &  4.11  &  52.7 \\
		       8  &  1.00  & 1.1 & 6.36  &  6.09  &  5.45  &  81.2 \\
		       \hline \\   
		\end{tabular}
		\label{tab:lambda}
\end{table}

\section{CTL during normal zone propagation}
We now investigate in details the behavior of the CTL when a quench occurs in HTS tapes. As in the static case, the CTL corresponds to the characteristic length required for the current to transfer from the superconductor to the silver stabilizer surrounding the NZ or vice-versa.
However, the dynamic CTL cannot be determined analytically: the highly nonlinear coupling between the electric and the thermal problems makes it too difficult. Therefore, we need to perform numerical simulations for this part of the analysis.

\subsection{Numerical simulations: finite element models} 
Two finite element (FE) models developed in COMSOL Multiphysics 4.3b were used in this paper. The first one (FE1) simulates the complete dynamic of quench propagation in HTS tapes, whereas the second one (FE2) simulates only the static surface potential on top of the silver layer (no heat equation). The two models are described below. 

\subsubsection{FE1 model: simulation of quench in HTS tapes}
The FE1 model consists of a 3-D finite element electro-thermal model that simulates the dynamics of quench propagation in HTS tapes, as described in details in~\cite{Lacroix2017,Fournier-Lupien2018a}. The model uses the Joule heating module in COMSOL 4.3b, which solves simultaneously the heat equation and the current continuity equation, which are strongly coupled. A NZ is initiated by a local heat pulse at one extremity of the tape. The geometry of the simulated tape is that of a classic CFD architecture~\cite{Lacroix2014a}, exactly as in figure~\ref{fig:fig1}, but without the groove. Post-processing on the results allowed extracting $\ld$, i.e~the dynamic CTL.

The $E-J$ characteristic of REBaCuO was modeled using an empirical temperature dependent power law $E(J,T)=E_0\left(J/J_c(T)\right)^{n(T)}$ in the flux creep and flux flow regimes~\cite{Brandt1996}, where $E_0=1~\mu$V/cm (electric field criterion at 77~K).
The transition from the superconducting state to the normal state was modeled by putting the conductivity of the superconducting and the normal state in parallel, such as two resistance in parallel in an electrical circuit. Once again, this is presented in details in~\cite{Lacroix2017,Fournier-Lupien2018a}.

In this paper, all the simulations performed with the FE1 model used the following parameters: i) a total current injected in the tape $I=0.9I_{c}$, with $I_{c}=100$~A at 77~K, ii) a critical temperature $T_c=90$~K, iii) an initial temperature of 77~K, iv) a tape width $w=12$~mm, v) interfacial resistances $R_f= 10^{3}~\mu\Omega.$m$^2$ and $R_0= 10^{-1}~\mu\Omega.$m$^2$, and vi) a silver layer thickness $d_{\textrm{Ag}}=1~\mu$m. The only parameter that was varied during the simulations was $w_f$, i.e.~the width of the CFD layer.

\subsubsection{FE2 model: simulation of surface potential}
The purpose of the FE2 model is to reproduce the experimental results presented in section~\ref{Exp_CTL} and to investigate the effect of the non-superconducting region (groove or NZ) on $\ls$, the static CTL. The FE2 model is a 3-D finite element model as the FE1 model, but it solves only the current continuity equation, not the heat equation. The temperature in the tape is therefore constant all along the simulation, and the output is the electric potential $V(x,y,z)$ in the tape. The model uses the Electric currents module in COMSOL 4.3b. For the simulations, the geometry presented in figure~\ref{fig:fig1}(b) was implemented, including the groove. The same geometric parameters as in the FE1 model were used with the FE2 model, except for the value of the transport current, which was $I=100$~mA.

\subsection{Simulation of the normal zone propagation}
\begin{figure}[tb] 
	\centering	
	\includegraphics[width=\columnwidth, trim={0cm 0cm 0cm 0cm},clip]{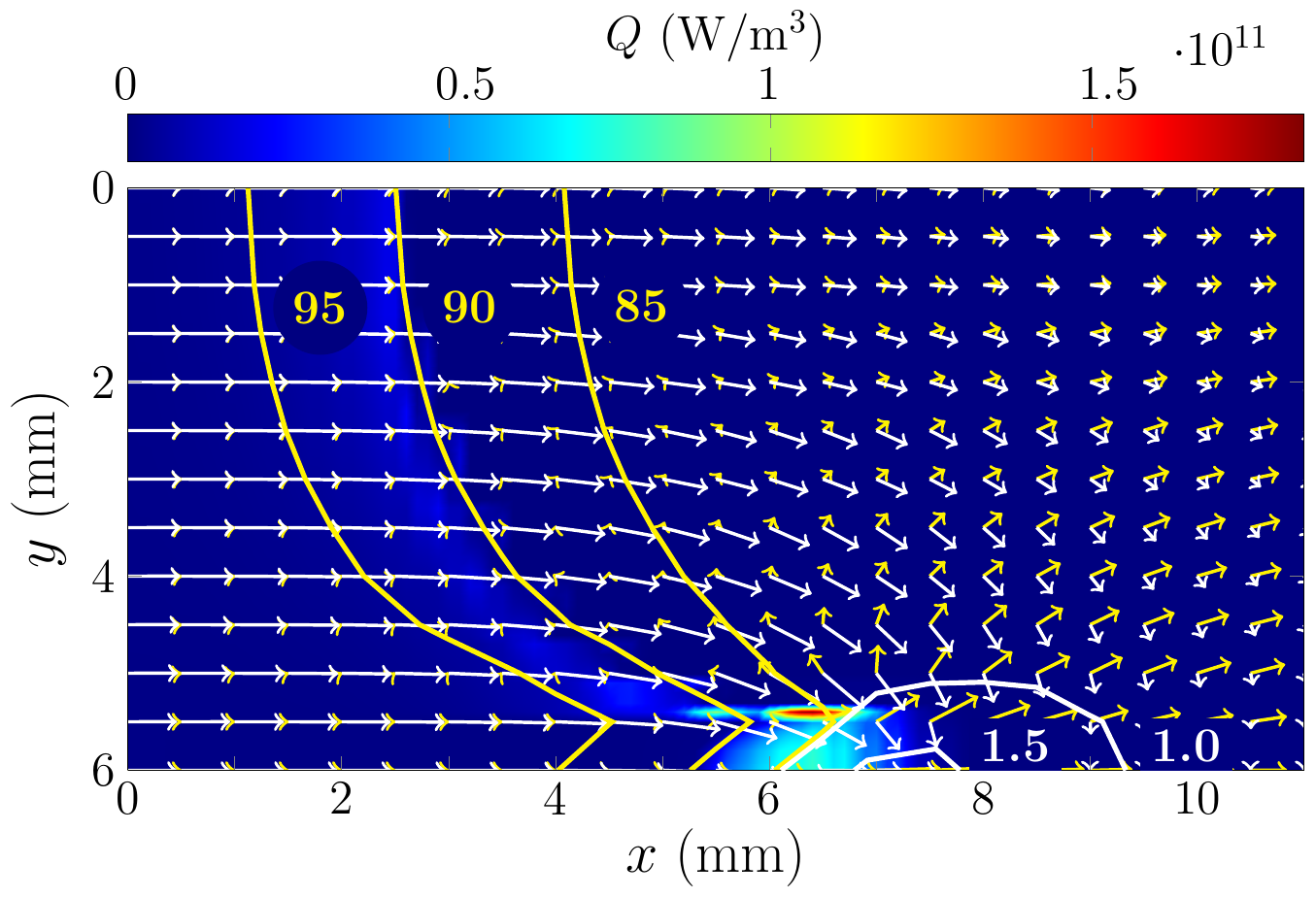} 
	\caption{
Snapshot of the simulation results for a quench propagation simulation performed with the FE1 model, for a HTS tape with CFD architecture and $f = 0.9$. Only a half-width of the tape is shown. See text for a detailed description of the figure.
}
	\label{fig:fig5-new}
\end{figure}

Figure~\ref{fig:fig5-new} shows a typical simulation output in the $x-y$ plane obtained with the FE1 model. In the simulation, the NZ was generated at $x=0$ and propagated in the $+x$ direction. The colormap represents the heat generation $Q$ (W/m$^3$) in the superconductor layer. The arrows represent the local current density in the superconductor (yellow) and in the silver stabilizer (white). The net current flows in the $+x$ direction. The isolines represent the superconductor temperature in K (yellow) and the magnitude of the current density normalized by the superconductor critical current density at 77~K (white), i.e. $J/J_c(77~\textrm{K})$.

Since the critical temperature $T_c=90$~K in the superconductor model, the entire region on the left side of the $T=90$~K isoline corresponds to the NZ. We also observe that the temperature isolines are curved, not straight.
These temperature profiles are typical for the CFD architectures and have already been observed experimentally~\cite{Fournier-Lupien2018a}. The curvature results from a current concentration on the edges of the tape, itself increasing the power density in the superconductor and the local heat generation. Indeed, we clearly see that the current density reaches values larger that $J_c$ in the superconductor, approximately 1 to 2~mm in front of the rightmost point of the NZ. This peculiar concentration, sometimes called ``current crowding'',  plays a key role in the determination of the CTL.

\subsection{Determination of the CTL from simulations}
\begin{figure}[t] 
	\centering	
	\includegraphics[width=\columnwidth, trim={0cm 0cm 0cm 0cm},clip]{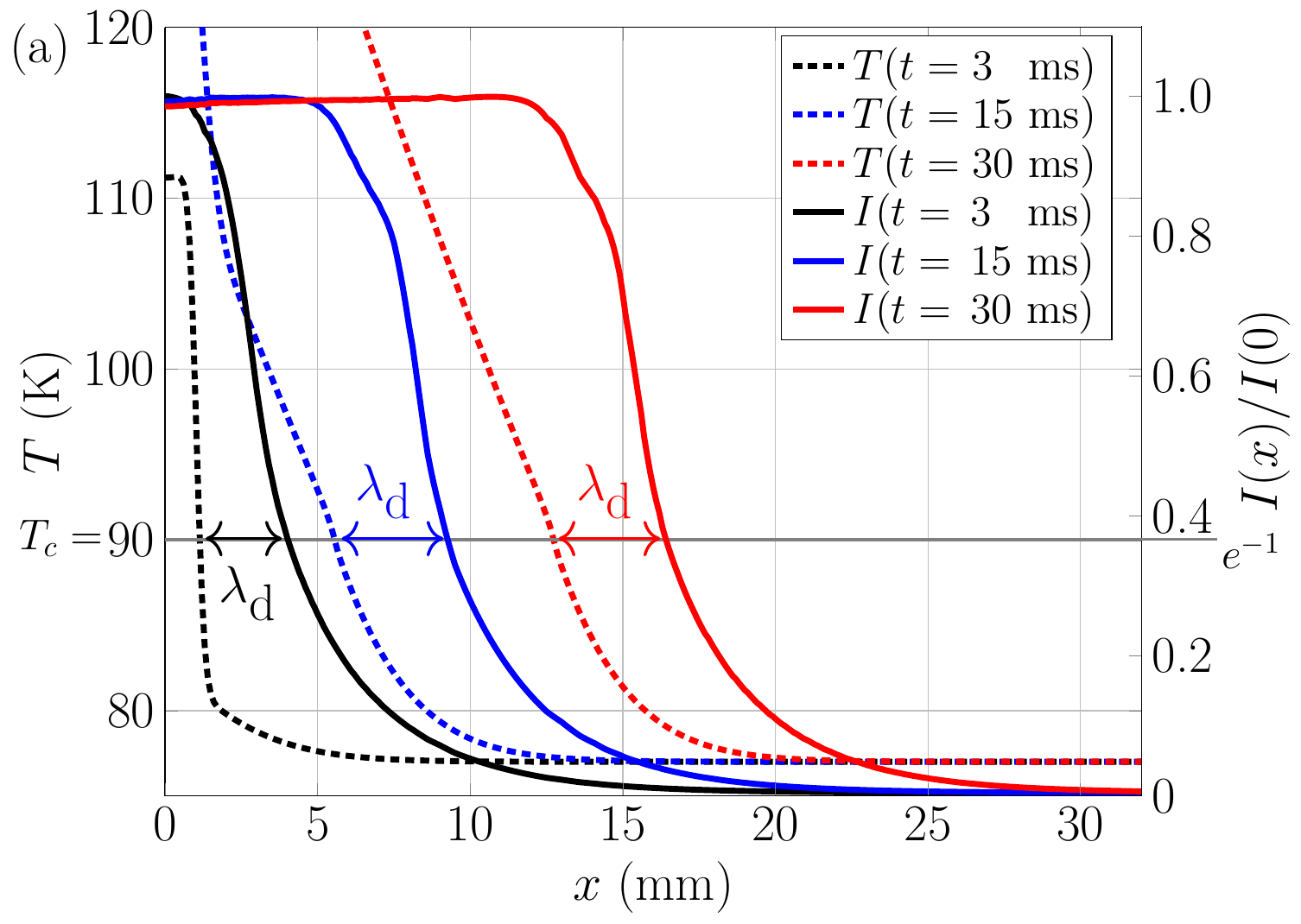}	
	\includegraphics[width=\columnwidth, trim={0cm 0cm 0cm 0cm},clip]{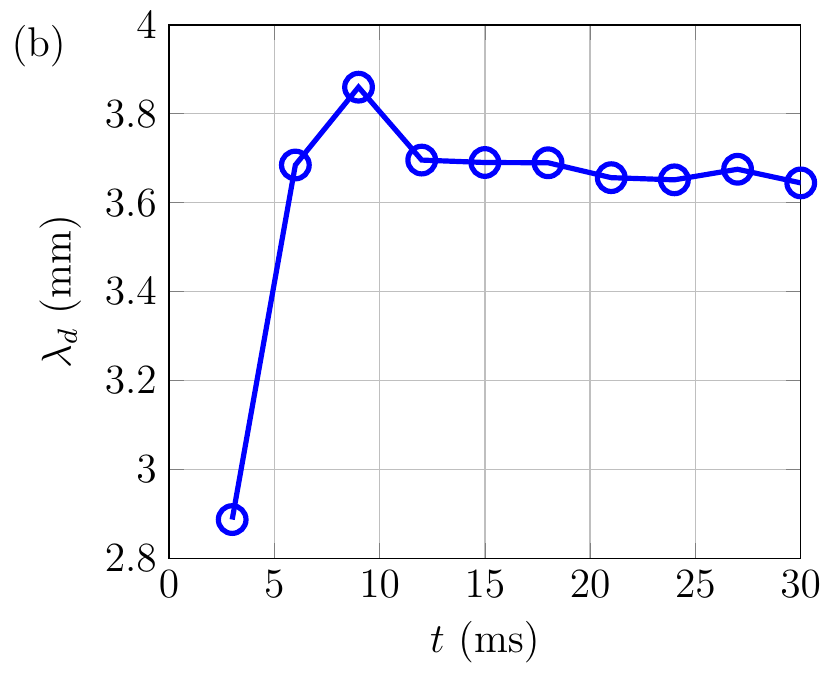}
	\caption{ (a) Simulation of a quench propagation (FE1 model) using the physical parameters of sample 5 from table~\ref{tab:optimi}. The NZ is initiated at $x=0$ and propagates along the $+x$ direction. The critical temperature $T_c=90$~K on the left axis is aligned with the current criterion $I(x)/I(0)=e^{-1}$. (b) Plot of the simulated CTL $\ldi$ extracted over time, as the NZ propagates. }
	\label{fig:fig6-new}
\end{figure}

We used the same definition as in section~\ref{CTL_exp} to extract the CTL from the finite element simulations, i.e. $I(\ld) / I(0) = e^{-1}$. Note that we used the notation $\ld$ in this case (dynamic CTL), since the current transfert region moves dynamically over time as the NZ expands. The boundary of the NZ is defined as the $T=T_c=90$~K isoline in the middle of the tape, i.e.~on the $y=0$ axis in figure~\ref{fig:fig5-new}.
Figure~\ref{fig:fig6-new}(a) illustrates how we used this definition to deduce $\ld$ at three different times of the simulation, namely 3~ms, 15~ms and 30~ms. We clearly see that the $T=90$~K point moves to the right as time increases, showing that the NZ expands. The value of $\ld$ is thus simply obtained by calculating the distance between the NZ boundary and the point where the normalized current has decreased to $e^{-1}$.

By repeating this process at many time steps, we obtained the evolution of $\ld$ as a function of time, illustrated in~\ref{fig:fig6-new}(b). We observe that it takes about 12~ms before $\ld$ reaches a steady state value.
For the purpose of our analysis, we disregard the transient part and consider only the steady-state (constant) value of $\ld$.

\begin{figure}[t] 
	\centering	
	\includegraphics[width=\columnwidth, trim={0cm 0cm 0cm 0cm},clip]{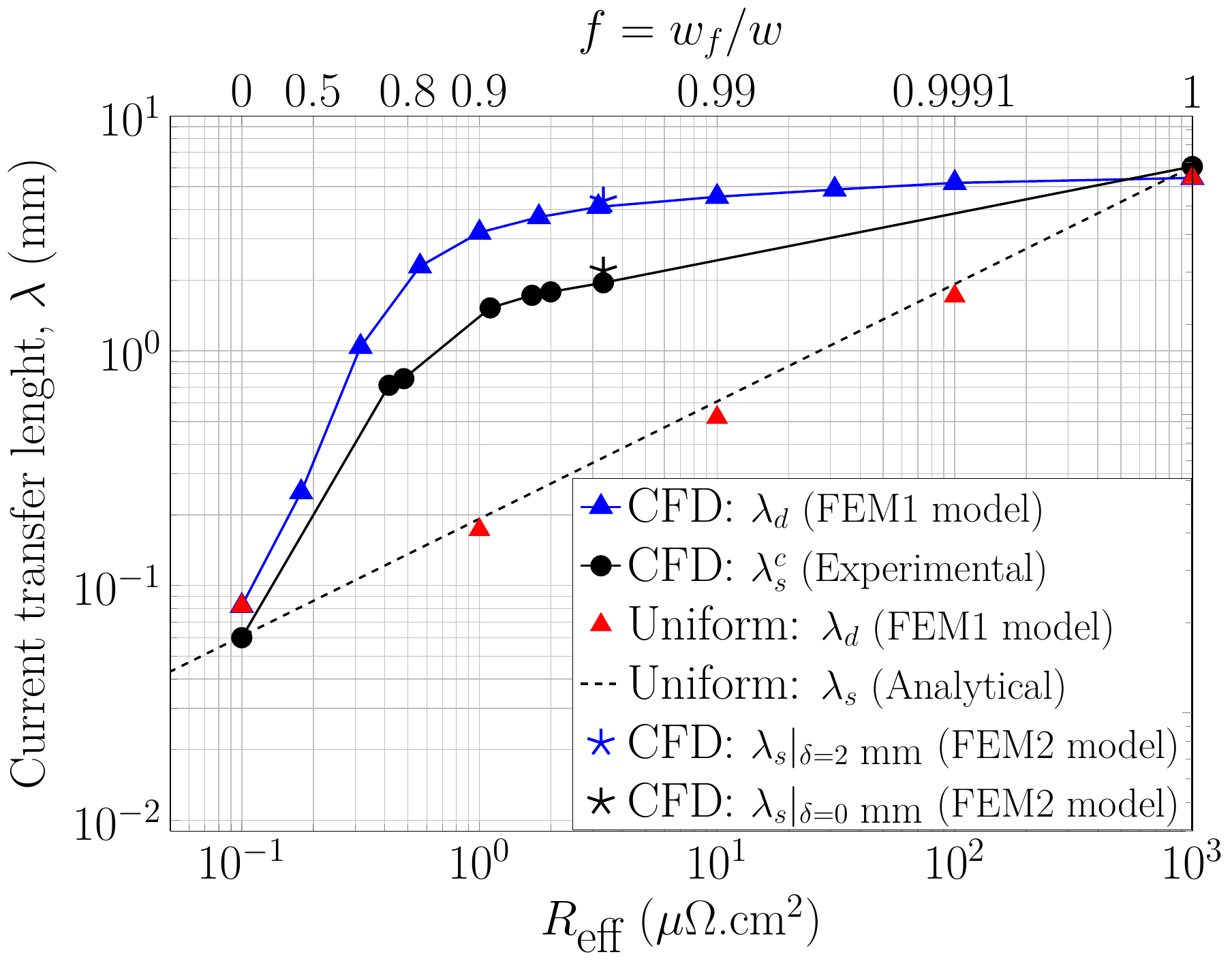}	
	\caption{Comparison of the static ($\ls$) and dynamic ($\ld$) CTL values for the CFD and uniform HTS tape architectures, respectively. All dynamic CTL values were obtained by simulation. The static CTL values for the CFD architecture were measured experimentally. The star symbols are discussed in section \ref{section:shapeNZ}.
}
	\label{fig:fig7}
\end{figure}

Many such simulations were performed to determine $\ld$ with the FE1 model for the CFD architecture, for values of $f$ ranging between 0 and 1, with $R_0$ and $R_f$ having constant values of $10^{-1}$ and $10^3$~$\Omega$.cm$^2$, respectively. The results are presented in figure~\ref{fig:fig7} (solid blue triangles). Numerical values of $\ld$ corresponding to the specific $f$ values of our measured samples are also reported in table~\ref{tab:lambda}, together with the simulated NZPV values. We clearly see in the figure how $\ld$ increases rapidly with $f$ up to $\approx 0.8-0.9$, and then remains quite constant. This explains why we usually chose a coverage fraction $f$ between 0.8 and 0.9 for the CFD layer.

We also used the FE1 model to calculate the $\ld$ values of the uniform architecture (red triangles), which can be seen as a CFD architecture with $f=1$ but with a variable $R_f$ value, here between $10^{-1}$ to $10^{3}~\mu \Omega.$cm$^2$. The simplest way to plot the two architectures on the same $x$ axis is to define the effective interfacial resistance of the tape as

\begin{equation}
	R_{\textrm{eff}} = \left( \frac{f}{R_f} + \frac{1-f}{R_0} \right)^{-1}.
\end{equation}
For the uniform architecture, the decrease of the current in the stabilizer is exponential, i.e.~$I(x)=e^{-x/\ls}$, with $\ls = (R_{\textrm{eff}}\, d_{\textrm{Ag}} / \rho_{\textrm{Ag}} )^{1/2}$. This analytical value of $\ls$ was added in figure~\ref{fig:fig7} (black dotted line) for the sake of validation of the numerical simulations (red triangles). We see that $\ls$ obtained with the analytical solution and $\ld$ obtained numerically with the FE2 model are essentially equivalent, as expected.

Finally, we also added in figure~\ref{fig:fig7} the measured CTL values $\lsc$ (the corrected values) reported in table~\ref{tab:lambda} (black circles) to support the discussion presented in next section. Note that these experimental values match perfectly the analytical values of $\ls$ for $f=0$ and $f=1$, which gives us confidence in our experimental procedure.

\subsection{Discussion about static and dynamic CTL}

We observe in figure~\ref{fig:fig7} that the CTL always increases monotonically as $R_\textrm{eff}$ increases, or alternatively, as $f$ increases for the CFD architecture. We also observe that the CFD architecture leads to larger values of both $\ld$ and $\ls$ with respect to the uniform architecture, except for $f=0$ or $f=1$, in which the two architectures are equivalent. However, the fundamental difference between these architectures is that $\ld\approx\ls$ for the uniform architecture, whereas $\ld>\ls$ for the CFD architecture. We develop the reasons below, after recalling the difference between the two types of CTL.

The static CTL $\ls$ corresponds to the CTL value when no heat generation is present. This is what we measured, for instance, in section~\ref{section:experiment} of this paper when we used a very small current in samples where a groove had been voluntarily added to break the continuity of the superconducting path. We can also obtain $\ls$ by numerical simulation if we solve solely the electrical problem and ignore the heat generation, as the FE2 model described above does. Note that the notion of static CTL does not exclude the existence of current crowding around the non-superconducting region.

The dynamic CTL $\ld$ corresponds to the CTL value when we consider thermal effects, and especially the NZ propagation. For observing $\ld$, it is necessary to apply currents higher than the local $I_c$, or at least higher than the weakest superconducting region of the tape, to ensure sufficient heating occurs. So far, we have not found a straightforward way to determine $\ld$ experimentally, so we always had to compute it by numerical simulation (FE1 model in this paper).

Based on the above definitions, since $\ld$ and $\ls$ are essentially equal for the uniform architecture, it means that the CTL of the uniform architecture is not affected by the thermal dynamics of the NZ propagation. Conversely, for the CFD architecture, $\ld$ is systematically larger than $\ls$, thus the CTL is affected by the thermal dynamics of the NZ. We shall see in next section that the condition required for $\ld$ and $\ls$ to be different is to introduce curvature in the shape of the NZ.


\subsection{Shape of the normal zone vs. CTL}
\label{section:shapeNZ}

It was shown in figure~\ref{fig:fig5-new} (yellow isolines) that the boundary of the NZ is curved for the CFD architecture.
This was in fact known since the original paper on the CFD concept (see figure~4 in~\cite{Lacroix2014b}), but the impact on the CTL had never been properly explained so far. In the same paper, it was also shown that the boundary of the NZ for the uniform architecture is a straight line rather than a curve~\cite{Lacroix2014b}.

\begin{figure}[tb] 
	\centering	
	\includegraphics[width=\columnwidth, trim={0cm 0cm 0cm 0cm},clip]{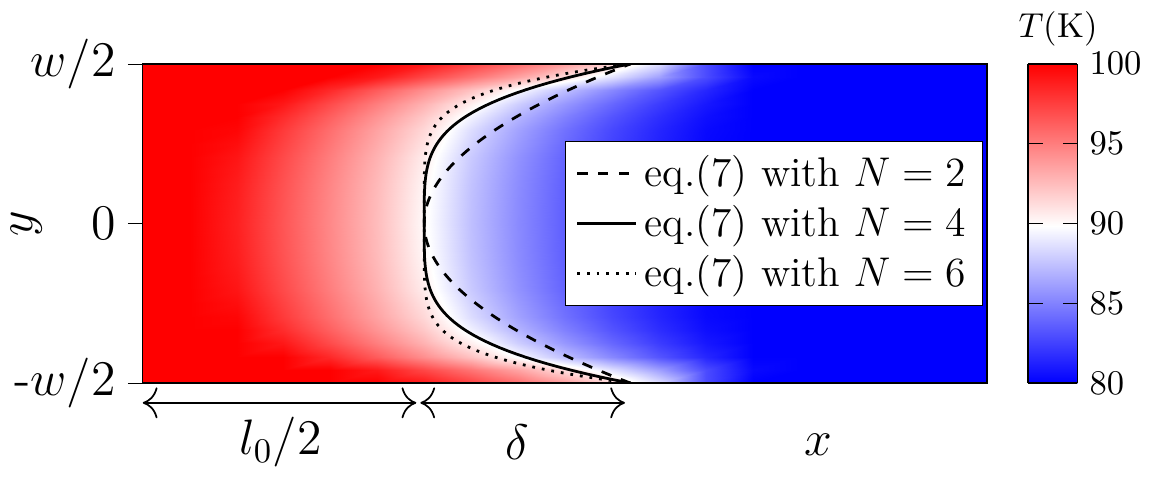} 
	\caption{ Equation~(\ref{eq:xu}) superimposed on a temperature map obtained with the FE1 model, during a quench propagation on a CFD HTS tape with $f=0.97$. The value of $\delta$ used for the three cases shown was 2~mm, and the final value selected for $N$ was 4.
}
	\label{fig:fig8}
\end{figure}

To analyze the effect of the NZ shape on the CTL, we used the FE2 model (3-D electric potential, no thermal equation) to compute the static CTL on the sample geometry presented in figure~\ref{fig:fig1}, for different shapes of the NZ. Note that this model solves the same equations as the SA model does, but with arbitrary shapes of NZ, which the SA model cannot do.
The following function was used to model the curved boundary of the NZ: 
\begin{equation} \label{eq:xu}
        x = l_0/2 + \delta \left( \frac{y}{w/2} \right) ^N \,,
\end{equation}
where $\delta$ and $N$ are tuning parameters used to produced the curved boundary. If $\delta=0$, then $x=l_0/2$, i.e.~we have a rectangular NZ (or a groove) whose width is $l_0$. If $\delta>0$, we have a rectangular NZ augmented by an additional curved region, as illustrated in figure~\ref{fig:fig8}. Note that this trick allows emulating in a purely geometric manner the curvature of the normal zone that normally arises from the thermal problem.

To find the values of $\delta$ and $N$ that best fit our simulation results, we superimposed equation~(\ref{eq:xu}) to the temperature map obtained with the FE1 model for a CFD tape with $f=0.97$ (see figure~\ref{fig:fig8}) and found that $\delta=2$~mm and $N=4$ were good choices. We then ran multiple simulations with the FE2 model for $N=4$ and various values of $\delta$ between 0~mm (rectangular NZ) and 3~mm (NZ with curved boundary). The simulations were run for three different CFD tapes ($f$ values of 0.79, 0.9 and 0.97, respectively).

\begin{figure}[tb] 
	\centering	
	\includegraphics[width=\columnwidth, trim={0cm 0cm 0cm 0cm},clip]{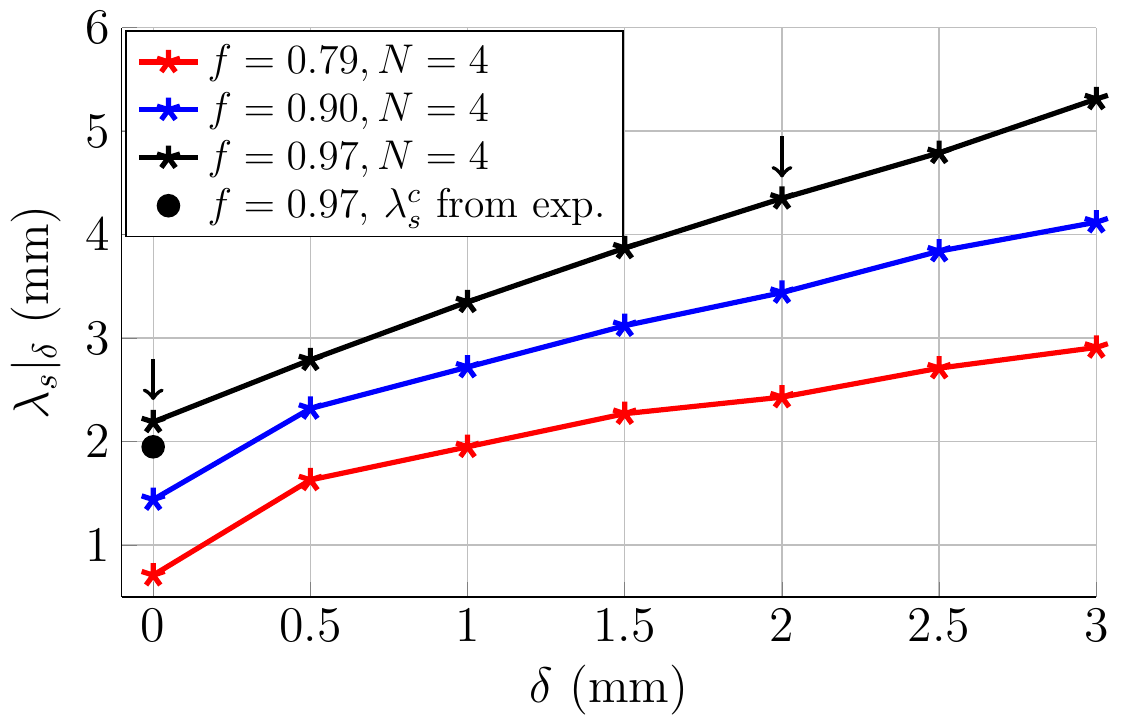} 
	\caption{
Static CTL $\ls$ calculated for CFD tapes with different $f$ values and different values of $\delta$, i.e. the width of curved region in the NZ, defined in~(\ref{eq:xu}). The arrows identify two points on the black curve that have been reported in figure~\ref{fig:fig7}.
}
	\label{fig:fig9}
\end{figure}

The static CTL values obtained, which we shall note $\ls |_\delta$ to emphasize the fact that we voluntarily modified the shape of the NZ boundary, were plotted in figure~\ref{fig:fig9} and show an approximately linear increase with the width $\delta$ of the curved region of the NZ. Remember that in reality, the curvature results from the existence of a non-uniform interfacial resistance between the superconductor and silver layers (because of the CFD layer), which induces 3-D current flow paths and current crowding (geometric effect), which in turn induces heat generation if the current is high enough (thermal effect). All these effects must be present to observe the NZ curvature.

We also added in figure~\ref{fig:fig9} the experimental value $\ls^c$ (black circle) for sample~7 ($f=0.97$), which was expected to fall on the black curve when $\delta=0$ (the groove then acts as the NZ). The point actually falls $\approx 10\%$ below the curve, but this is considered within the experimental error and thus confirms the validity of the simulations.


%

To complete the demonstration of the influence of the shape of the NZ on the CTL, we reported in figure~\ref{fig:fig7} two specific values of the calculated $\ls |_\delta$ in figure~\ref{fig:fig9} (star markers), namely $\ls |_{(\delta=0~\textrm{mm})}$ and $\ls |_{(\delta=2~\textrm{mm})}$, both corresponding to the case $f=0.97$ and $N=4$. We clearly see in~figure~\ref{fig:fig7} that the $\ls |_{(\delta=0~\textrm{mm})}$ point (black star) falls very close to the experimental static CTL curve (solid black circles), and that the $\ls |_{(\delta=2~\textrm{mm})}$ point (blue star) falls very close to the dynamic CTL value (solid blue triangles). Since both points were calculated with the FE2 model, providing by definition a static CTL value, it is now clear that the primary factor affecting the CTL is the geometric shape of the NZ.

This raises an interesting point about the definition of the CTL. The classic definition of the CTL is that of $\ls$, the static CTL, which does not consider thermal effects~\cite{Ekin1978,Stenvall2007}. However, we have proved that we can retrieve the dynamic CTL $\ld$ from the definition of $\ls$ by considering the shape of the NZ when calculating the static CTL (noted $\ls |_\delta$ above). By doing this, we have decoupled the geometric and thermal effects in the determination of the CTL. When considering the uniform tape architecture, this decoupling is pointless since the NZ is rectangular (or nearly rectangular) during a quench propagation, and $\ld \approx \ls$, which means that the heat propagation is essentially 1-D and has no impact on the CTL. However, with the CFD architecture, the strong current crowding on the edges of the tape (3-D behavior) provides the geometric effect required to trigger inhomogeneous heating (thermal effect), which in turn produces the curved iso-temperature lines of figure~\ref{fig:fig5-new}, thus the curved normal zone boundary. In this case, we have $\ld > \ls$, where $\ls$ is the CTL resulting from the sole geometric effect (rectangular NZ), and $\ld$ is the CTL resulting from the combination of the geometric and thermal effects (NZ with curved boundaries).

Finally, this also explains why $\ls$ in CFD tapes is systematically larger than $\ls$ in uniform tapes  for the same values of $R_\textrm{eff}$ (see black circles vs red triangles in figure~\ref{fig:fig7}). Since no thermal contribution enters in the $\ls$ definition, the difference come exclusively from the geometric effect (current crowding) present in CFD tapes, and not present in uniform tapes. Note that this conclusion applies without the need to invoke any curvature in the shape of the NZ, as shown in figure~\ref{fig:fig7} for tapes with rectangular grooves acting as the NZ.

At this point, if we focus only on the CFD architecture, there remains three important questions to answer, namely: 1) How are $\ld$ and $\ls$ related to each other? 2) How is the NZPV related to $\ls$ and $\ld$? 3) Is it worth the extra work of determining $\ld$? The next section provides answers to these questions.

\section{Further considerations about $\ls$ and $\ld$ in CFD tapes}

\begin{figure}[t] 
	\centering	
	\includegraphics[width=\columnwidth, trim={0cm 0cm 0cm 0cm},clip]{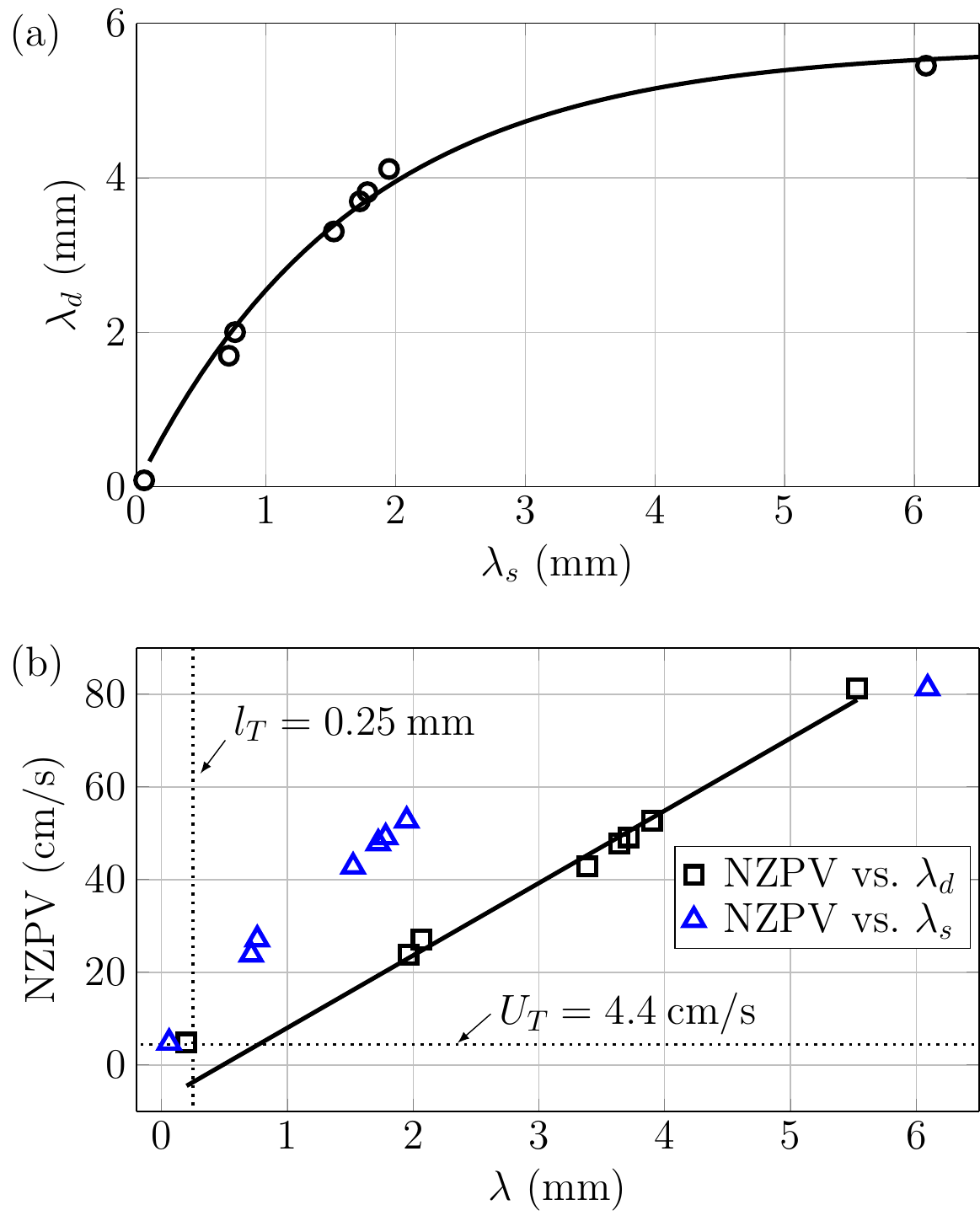} 
\caption{
(a) Dynamic ($\ld$) vs.~static ($\ls$) CTL values for the eight CFD samples considered in this paper (circles), and tentative exponential fit of the observed relationship (solid line), i.e. $\ld=a(1-e^{-b\ls})$. (b) Normal zone propagation velocity (NZPV) extracted from the simulations vs.~the dynamic (square) and static (triangle) CTL for the eight CFD samples considered in this paper, revealing a nearly perfect linear fit when plotted against $\ld$ for $\lambda>l_T$ (solid line, see text for explanation).
}
	\label{fig:fig10}
\end{figure}

Figure~\ref{fig:fig10}(a) shows a plot of $\ld$ (simulated) versus $\ls$ (measured) for the eight CFD samples considered in this paper. The points can be well fitted by an exponential relationship such as $\ld=a(1-e^{-b\ls})$, with $a=5.683$~mm and $b=0.594$~mm$^{-1}$. Although this is a simple relationship, it is not clear from this single experiment how the $a$ and $b$ parameters would vary with the numerous geometric parameters defining a CFD tape architecture, but it would be an interesting avenue to explore.

In figure~\ref{fig:fig10}(b), we plotted the calculated NZPV values as a function of the calculated $\ld$ values (all values were taken directly from table~\ref{tab:lambda}), for the eight CFD samples investigated. Both quantities were calculated with the FE1 model, but the NZPV could have been measured experimentally. It is striking from this figure that the NZPV is directly proportional to $\ld$, the dynamic CTL. There is no such nice relationship between the NZPV and the measured $\ls$ values (blue triangles).

When looking more closely at the NZPV vs.~$\ld$ plot, we see that only the first point of the curve does not fit well with the linear relationship. This point corresponds to $f=0$, i.e.~a tape without CFD. This does not come as a surprise, since Levin showed in~\cite{Levin2010a} that the NZPV is proportional to the CTL only when the CTL is larger than the thermal diffusion length $l_T$. Using Levin's equations in~\cite{Levin2010a}, we find $l_T=0.25$~mm (a constant) for the HTS tapes considered in this work. This implies that the NZPV varies linearly only for $\lambda_d>l_T$, which is consistent with what we see in figure~\ref{fig:fig10}(b). For $\lambda_d<l_T$, the NZPV converges towards a constant value noted $U_T$, equal to 4.4~cm/s in the current case~\cite{Levin2010a}. Therefore, the apparent outlier data point perfectly fit with the theory.

As a final comment, the clear proportionality observed between the NZPV and $\ld$ suggests that the dynamic CTL is more adequate than the static CTL $\ls$ for studying HTS tapes with CFD-like architectures. The dynamic CTL should thus probably be considered as the true general definition of the CTL. Therefore, to answer our last question, it seems that calculating $\ld$ of CFD tapes is worth the effort. In addition, the perfect linear relationship found between the NZPV and $\ld$ might provide an experimental way to relate the measurable static CTL to the ``more-difficult-to-measure'' dynamic CTL in future work.

\section{Conclusion}
In this work, we used a mixed experimental and numerical simulation methodology to show for the first time that the current transfer length (CTL) in 2G HTS tapes result from two distinct effects, namely i) a geometric effect, and ii) a thermal effect. The geometric effect is caused by a non-uniform interfacial resistance between the superconductor and the stabilizer layers, and results in non-uniform current transfer pattern (2-D effect) and current crowding. The thermal effect consists in inhomogeneous heat generation during the current transfer, with higher heat generation in the higher current density regions. The thermal effect, if strong enough, generates curvature in the normal zone (NZ) boundaries, which in turn affects the current flow paths. Two conditions are required to observe thermal effects: i) the transport current must be high enough, and ii) the geometric effect (current crowding) must be present.

Based on these observations, we could defined two types of CTL, namely i) the static CTL $\ls$, which corresponds to a CTL that considers solely the geometric effect (no heat propagation), and ii) the dynamic CTL $\ld$, which takes into account both the geometric and thermal effects. Note that $\ls$ corresponds to the classical definition of the CTL. For uniform HTS tape architectures, which correspond essentially to nowadays commercial tapes, $\ls$ and $\ld$ are equal, since no significant geometric effect is present. However, for the CFD tape architecture, there is a clear distinction between $\ls$ and $\ld$, with $\ld$ being systematically larger than $\ls$. Furthermore, because of the geometric effect present in the CFD architecture, $\ls$ in CFD tapes is systematically larger than $\ls$ in uniform tapes.

As a final observation, we confirmed that the normal zone propagation velocity (NZPV) in HTS tapes is proportional to $\ld$, not to $\ls$, which strongly suggests that $\ld$ is a more general definition of the CTL than $\ls$. Therefore, in order to perform a fair comparison between any type of HTS tape architecture, in particular CFD-like architectures~\cite{Lacroix2014b,fournier2018}, the dynamic CTL $\ld$ should always be used.

\section{Acknowledgement}
The authors would like to thank Marc-Antoine Huet for the design and realization of the multiplexer system used with the pogo pin arrays. They also thank Mohammed Amine Drai for the design of the 6-layer PCB on which the arrays were soldered. Finally, J.-H. Fournier-Lupien and F. Sirois gratefully thank Prof.~Mathias Noe for hosting them at KIT in 2015-2016, where this work was initiated.
 
\bibliography{ library }

\begin{thebibliography}{38}%
\makeatletter
\providecommand \@ifxundefined [1]{%
 \@ifx{#1\undefined}
}%
\providecommand \@ifnum [1]{%
 \ifnum #1\expandafter \@firstoftwo
 \else \expandafter \@secondoftwo
 \fi
}%
\providecommand \@ifx [1]{%
 \ifx #1\expandafter \@firstoftwo
 \else \expandafter \@secondoftwo
 \fi
}%
\providecommand \natexlab [1]{#1}%
\providecommand \enquote  [1]{``#1''}%
\providecommand \bibnamefont  [1]{#1}%
\providecommand \bibfnamefont [1]{#1}%
\providecommand \citenamefont [1]{#1}%
\providecommand \href@noop [0]{\@secondoftwo}%
\providecommand \href [0]{\begingroup \@sanitize@url \@href}%
\providecommand \@href[1]{\@@startlink{#1}\@@href}%
\providecommand \@@href[1]{\endgroup#1\@@endlink}%
\providecommand \@sanitize@url [0]{\catcode `\\12\catcode `\$12\catcode
  `\&12\catcode `\#12\catcode `\^12\catcode `\_12\catcode `\%12\relax}%
\providecommand \@@startlink[1]{}%
\providecommand \@@endlink[0]{}%
\providecommand \url  [0]{\begingroup\@sanitize@url \@url }%
\providecommand \@url [1]{\endgroup\@href {#1}{\urlprefix }}%
\providecommand \urlprefix  [0]{URL }%
\providecommand \Eprint [0]{\href }%
\providecommand \doibase [0]{http://dx.doi.org/}%
\providecommand \selectlanguage [0]{\@gobble}%
\providecommand \bibinfo  [0]{\@secondoftwo}%
\providecommand \bibfield  [0]{\@secondoftwo}%
\providecommand \translation [1]{[#1]}%
\providecommand \BibitemOpen [0]{}%
\providecommand \bibitemStop [0]{}%
\providecommand \bibitemNoStop [0]{.\EOS\space}%
\providecommand \EOS [0]{\spacefactor3000\relax}%
\providecommand \BibitemShut  [1]{\csname bibitem#1\endcsname}%
\let\auto@bib@innerbib\@empty
\bibitem [{\citenamefont {Trociewitz}\ \emph {et~al.}(2011)\citenamefont
  {Trociewitz}, \citenamefont {Dalban-Canassy}, \citenamefont {Hannion},
  \citenamefont {Hilton}, \citenamefont {Jaroszynski}, \citenamefont {Noyes},
  \citenamefont {Viouchkov}, \citenamefont {Weijers},\ and\ \citenamefont
  {Larbalestier}}]{Trociewitz2011}%
  \BibitemOpen
  \bibfield  {author} {\bibinfo {author} {\bibfnamefont {U.~P.}\ \bibnamefont
  {Trociewitz}}, \bibinfo {author} {\bibfnamefont {M.}~\bibnamefont
  {Dalban-Canassy}}, \bibinfo {author} {\bibfnamefont {M.}~\bibnamefont
  {Hannion}}, \bibinfo {author} {\bibfnamefont {D.~K.}\ \bibnamefont {Hilton}},
  \bibinfo {author} {\bibfnamefont {J.}~\bibnamefont {Jaroszynski}}, \bibinfo
  {author} {\bibfnamefont {P.}~\bibnamefont {Noyes}}, \bibinfo {author}
  {\bibfnamefont {Y.}~\bibnamefont {Viouchkov}}, \bibinfo {author}
  {\bibfnamefont {H.~W.}\ \bibnamefont {Weijers}}, \ and\ \bibinfo {author}
  {\bibfnamefont {D.~C.}\ \bibnamefont {Larbalestier}},\ }\href
  {https://aip.scitation.org/doi/full/10.1063/1.3662963} {\bibfield  {journal}
  {\bibinfo  {journal} {Applied Physics Letters}\ }\textbf {\bibinfo {volume}
  {99}},\ \bibinfo {pages} {202506} (\bibinfo {year} {2011})}\BibitemShut
  {NoStop}%
\bibitem [{\citenamefont {Senatore}\ \emph {et~al.}(2014)\citenamefont
  {Senatore}, \citenamefont {Alessandrini}, \citenamefont {Lucarelli},
  \citenamefont {Tediosi}, \citenamefont {Uglietti},\ and\ \citenamefont
  {Iwasa}}]{Senatore2014}%
  \BibitemOpen
  \bibfield  {author} {\bibinfo {author} {\bibfnamefont {C.}~\bibnamefont
  {Senatore}}, \bibinfo {author} {\bibfnamefont {M.}~\bibnamefont
  {Alessandrini}}, \bibinfo {author} {\bibfnamefont {A.}~\bibnamefont
  {Lucarelli}}, \bibinfo {author} {\bibfnamefont {R.}~\bibnamefont {Tediosi}},
  \bibinfo {author} {\bibfnamefont {D.}~\bibnamefont {Uglietti}}, \ and\
  \bibinfo {author} {\bibfnamefont {Y.}~\bibnamefont {Iwasa}},\ }\href
  {https://iopscience.iop.org/article/10.1088/0953-2048/27/10/103001/pdf}
  {\bibfield  {journal} {\bibinfo  {journal} {Superconductor Science and
  Technology}\ }\textbf {\bibinfo {volume} {27}},\ \bibinfo {pages} {103001}
  (\bibinfo {year} {2014})}\BibitemShut {NoStop}%
\bibitem [{\citenamefont {Maeda}\ and\ \citenamefont
  {Yanagisawa}(2019)}]{Maeda2019}%
  \BibitemOpen
  \bibfield  {author} {\bibinfo {author} {\bibfnamefont {H.}~\bibnamefont
  {Maeda}}\ and\ \bibinfo {author} {\bibfnamefont {Y.}~\bibnamefont
  {Yanagisawa}},\ }\href {https://doi.org/10.1016/j.jmr.2019.07.011} {\bibfield
   {journal} {\bibinfo  {journal} {Journal of Magnetic Resonance}\ }\textbf
  {\bibinfo {volume} {306}},\ \bibinfo {pages} {80} (\bibinfo {year}
  {2019})}\BibitemShut {NoStop}%
\bibitem [{\citenamefont {Uglietti}(2019)}]{Uglietti2019}%
  \BibitemOpen
  \bibfield  {author} {\bibinfo {author} {\bibfnamefont {D.}~\bibnamefont
  {Uglietti}},\ }\href
  {https://iopscience.iop.org/article/10.1088/1361-6668/ab06a2} {\bibfield
  {journal} {\bibinfo  {journal} {Superconductor Science and Technology}\
  }\textbf {\bibinfo {volume} {32}},\ \bibinfo {pages} {053001} (\bibinfo
  {year} {2019})}\BibitemShut {NoStop}%
\bibitem [{\citenamefont {Levin}\ \emph {et~al.}(2009)\citenamefont {Levin},
  \citenamefont {Novak},\ and\ \citenamefont {Barnes}}]{Levin2010a}%
  \BibitemOpen
  \bibfield  {author} {\bibinfo {author} {\bibfnamefont {G.}~\bibnamefont
  {Levin}}, \bibinfo {author} {\bibfnamefont {K.}~\bibnamefont {Novak}}, \ and\
  \bibinfo {author} {\bibfnamefont {P.}~\bibnamefont {Barnes}},\ }\href
  {https://iopscience.iop.org/article/10.1088/0953-2048/23/1/014021/meta}
  {\bibfield  {journal} {\bibinfo  {journal} {Superconductor Science and
  Technology}\ }\textbf {\bibinfo {volume} {23}},\ \bibinfo {pages} {014021}
  (\bibinfo {year} {2009})}\BibitemShut {NoStop}%
\bibitem [{\citenamefont {Lacroix}\ \emph {et~al.}(2014)\citenamefont
  {Lacroix}, \citenamefont {Lapierre}, \citenamefont {Coulombe},\ and\
  \citenamefont {Sirois}}]{Lacroix2014a}%
  \BibitemOpen
  \bibfield  {author} {\bibinfo {author} {\bibfnamefont {C.}~\bibnamefont
  {Lacroix}}, \bibinfo {author} {\bibfnamefont {Y.}~\bibnamefont {Lapierre}},
  \bibinfo {author} {\bibfnamefont {J.}~\bibnamefont {Coulombe}}, \ and\
  \bibinfo {author} {\bibfnamefont {F.}~\bibnamefont {Sirois}},\ }\href
  {https://iopscience.iop.org/article/10.1088/0953-2048/27/5/055013/meta}
  {\bibfield  {journal} {\bibinfo  {journal} {Superconductor Science and
  technology}\ }\textbf {\bibinfo {volume} {27}},\ \bibinfo {pages} {055013}
  (\bibinfo {year} {2014})}\BibitemShut {NoStop}%
\bibitem [{\citenamefont {Levin}\ \emph {et~al.}(2007)\citenamefont {Levin},
  \citenamefont {Barnes},\ and\ \citenamefont {Bulmer}}]{Levin2007b}%
  \BibitemOpen
  \bibfield  {author} {\bibinfo {author} {\bibfnamefont {G.~A.}\ \bibnamefont
  {Levin}}, \bibinfo {author} {\bibfnamefont {P.~N.}\ \bibnamefont {Barnes}}, \
  and\ \bibinfo {author} {\bibfnamefont {J.~S.}\ \bibnamefont {Bulmer}},\
  }\href {https://iopscience.iop.org/article/10.1088/0953-2048/20/8/006}
  {\bibfield  {journal} {\bibinfo  {journal} {Superconductor Science and
  Technology}\ }\textbf {\bibinfo {volume} {20}},\ \bibinfo {pages} {757}
  (\bibinfo {year} {2007})}\BibitemShut {NoStop}%
\bibitem [{\citenamefont {Ekin}(1978)}]{Ekin1978}%
  \BibitemOpen
  \bibfield  {author} {\bibinfo {author} {\bibfnamefont {J.}~\bibnamefont
  {Ekin}},\ }\href {https://aip.scitation.org/doi/abs/10.1063/1.325245}
  {\bibfield  {journal} {\bibinfo  {journal} {Journal of Applied Physics}\
  }\textbf {\bibinfo {volume} {49}},\ \bibinfo {pages} {3406} (\bibinfo {year}
  {1978})}\BibitemShut {NoStop}%
\bibitem [{\citenamefont {Dhalle}\ \emph {et~al.}(1999)\citenamefont {Dhalle},
  \citenamefont {Porcar}, \citenamefont {Ivancevic}, \citenamefont {Polcari},
  \citenamefont {Huang}, \citenamefont {Witz},\ and\ \citenamefont
  {Flukiger}}]{Dhalle1999}%
  \BibitemOpen
  \bibfield  {author} {\bibinfo {author} {\bibfnamefont {M.}~\bibnamefont
  {Dhalle}}, \bibinfo {author} {\bibfnamefont {L.}~\bibnamefont {Porcar}},
  \bibinfo {author} {\bibfnamefont {M.}~\bibnamefont {Ivancevic}}, \bibinfo
  {author} {\bibfnamefont {A.}~\bibnamefont {Polcari}}, \bibinfo {author}
  {\bibfnamefont {Y.}~\bibnamefont {Huang}}, \bibinfo {author} {\bibfnamefont
  {G.}~\bibnamefont {Witz}}, \ and\ \bibinfo {author} {\bibfnamefont
  {R.}~\bibnamefont {Flukiger}},\ }\href
  {https://ieeexplore.ieee.org/abstract/document/783488?casa_token=59oxrOIVpokAAAAA:MuJuXp6fmBcoEAwkl6MqcGYHwcdz5KhI1jY8bkl1nNKexYb2wY7aIkCt7HFAt2-Cz0x8-iuI9nsy}
  {\bibfield  {journal} {\bibinfo  {journal} {IEEE transactions on applied
  superconductivity}\ }\textbf {\bibinfo {volume} {9}},\ \bibinfo {pages}
  {1093} (\bibinfo {year} {1999})}\BibitemShut {NoStop}%
\bibitem [{\citenamefont {Polak}\ \emph {et~al.}(1997)\citenamefont {Polak},
  \citenamefont {Zhang}, \citenamefont {Parrell}, \citenamefont {Cai},
  \citenamefont {Polyanskii}, \citenamefont {Hellstrom}, \citenamefont
  {Larbalestier},\ and\ \citenamefont {Majoros}}]{Polak1997}%
  \BibitemOpen
  \bibfield  {author} {\bibinfo {author} {\bibfnamefont {M.}~\bibnamefont
  {Polak}}, \bibinfo {author} {\bibfnamefont {W.}~\bibnamefont {Zhang}},
  \bibinfo {author} {\bibfnamefont {J.}~\bibnamefont {Parrell}}, \bibinfo
  {author} {\bibfnamefont {X.}~\bibnamefont {Cai}}, \bibinfo {author}
  {\bibfnamefont {A.}~\bibnamefont {Polyanskii}}, \bibinfo {author}
  {\bibfnamefont {E.}~\bibnamefont {Hellstrom}}, \bibinfo {author}
  {\bibfnamefont {D.}~\bibnamefont {Larbalestier}}, \ and\ \bibinfo {author}
  {\bibfnamefont {M.}~\bibnamefont {Majoros}},\ }\href
  {https://iopscience.iop.org/article/10.1088/0953-2048/10/10/007/meta}
  {\bibfield  {journal} {\bibinfo  {journal} {Superconductor Science and
  Technology}\ }\textbf {\bibinfo {volume} {10}},\ \bibinfo {pages} {769}
  (\bibinfo {year} {1997})}\BibitemShut {NoStop}%
\bibitem [{\citenamefont {Stenvall}\ \emph {et~al.}(2006)\citenamefont
  {Stenvall}, \citenamefont {Korpela}, \citenamefont {Lehtonen},\ and\
  \citenamefont {Mikkonen}}]{Stenvall2007}%
  \BibitemOpen
  \bibfield  {author} {\bibinfo {author} {\bibfnamefont {A.}~\bibnamefont
  {Stenvall}}, \bibinfo {author} {\bibfnamefont {A.}~\bibnamefont {Korpela}},
  \bibinfo {author} {\bibfnamefont {J.}~\bibnamefont {Lehtonen}}, \ and\
  \bibinfo {author} {\bibfnamefont {R.}~\bibnamefont {Mikkonen}},\ }\href
  {https://iopscience.iop.org/article/10.1088/0953-2048/20/1/017/meta}
  {\bibfield  {journal} {\bibinfo  {journal} {Superconductor Science and
  Technology}\ }\textbf {\bibinfo {volume} {20}},\ \bibinfo {pages} {92}
  (\bibinfo {year} {2006})}\BibitemShut {NoStop}%
\bibitem [{\citenamefont {Hol{\'u}bek}\ \emph {et~al.}(2007)\citenamefont
  {Hol{\'u}bek}, \citenamefont {Dhall{\'e}},\ and\ \citenamefont
  {Kov{\'a}{\v{c}}}}]{holubek2007}%
  \BibitemOpen
  \bibfield  {author} {\bibinfo {author} {\bibfnamefont {T.}~\bibnamefont
  {Hol{\'u}bek}}, \bibinfo {author} {\bibfnamefont {M.}~\bibnamefont
  {Dhall{\'e}}}, \ and\ \bibinfo {author} {\bibfnamefont {P.}~\bibnamefont
  {Kov{\'a}{\v{c}}}},\ }\href@noop {} {\bibfield  {journal} {\bibinfo
  {journal} {Superconductor Science and Technology}\ }\textbf {\bibinfo
  {volume} {20}},\ \bibinfo {pages} {123} (\bibinfo {year} {2007})}\BibitemShut
  {NoStop}%
\bibitem [{\citenamefont {Vinod}\ \emph {et~al.}(2010)\citenamefont {Vinod},
  \citenamefont {Varghese}, \citenamefont {Rahul}, \citenamefont {Devadas},
  \citenamefont {Thomas}, \citenamefont {Gurusamy}, \citenamefont {Kedia},
  \citenamefont {Pradhan},\ and\ \citenamefont {Syamaprasad}}]{vinod2010}%
  \BibitemOpen
  \bibfield  {author} {\bibinfo {author} {\bibfnamefont {K.}~\bibnamefont
  {Vinod}}, \bibinfo {author} {\bibfnamefont {N.}~\bibnamefont {Varghese}},
  \bibinfo {author} {\bibfnamefont {S.}~\bibnamefont {Rahul}}, \bibinfo
  {author} {\bibfnamefont {K.}~\bibnamefont {Devadas}}, \bibinfo {author}
  {\bibfnamefont {S.}~\bibnamefont {Thomas}}, \bibinfo {author} {\bibfnamefont
  {P.}~\bibnamefont {Gurusamy}}, \bibinfo {author} {\bibfnamefont
  {S.}~\bibnamefont {Kedia}}, \bibinfo {author} {\bibfnamefont
  {S.}~\bibnamefont {Pradhan}}, \ and\ \bibinfo {author} {\bibfnamefont
  {U.}~\bibnamefont {Syamaprasad}},\ }\href@noop {} {\bibfield  {journal}
  {\bibinfo  {journal} {Superconductor Science and Technology}\ }\textbf
  {\bibinfo {volume} {23}},\ \bibinfo {pages} {105002} (\bibinfo {year}
  {2010})}\BibitemShut {NoStop}%
\bibitem [{\citenamefont {Polak}\ \emph {et~al.}(2006)\citenamefont {Polak},
  \citenamefont {Barnes},\ and\ \citenamefont {Levin}}]{Polak2006}%
  \BibitemOpen
  \bibfield  {author} {\bibinfo {author} {\bibfnamefont {M.}~\bibnamefont
  {Polak}}, \bibinfo {author} {\bibfnamefont {P.}~\bibnamefont {Barnes}}, \
  and\ \bibinfo {author} {\bibfnamefont {G.}~\bibnamefont {Levin}},\ }\href
  {https://iopscience.iop.org/article/10.1088/0953-2048/19/8/022/meta}
  {\bibfield  {journal} {\bibinfo  {journal} {Superconductor Science and
  Technology}\ }\textbf {\bibinfo {volume} {19}},\ \bibinfo {pages} {817}
  (\bibinfo {year} {2006})}\BibitemShut {NoStop}%
\bibitem [{\citenamefont {Duckworth}\ \emph {et~al.}(2009)\citenamefont
  {Duckworth}, \citenamefont {List},\ and\ \citenamefont
  {Zhang}}]{Duckworth2009}%
  \BibitemOpen
  \bibfield  {author} {\bibinfo {author} {\bibfnamefont {R.~C.}\ \bibnamefont
  {Duckworth}}, \bibinfo {author} {\bibfnamefont {F.~A.}\ \bibnamefont {List}},
  \ and\ \bibinfo {author} {\bibfnamefont {Y.}~\bibnamefont {Zhang}},\ }\href
  {https://ieeexplore.ieee.org/abstract/document/4967857} {\bibfield  {journal}
  {\bibinfo  {journal} {IEEE transactions on applied superconductivity}\
  }\textbf {\bibinfo {volume} {19}},\ \bibinfo {pages} {3327} (\bibinfo {year}
  {2009})}\BibitemShut {NoStop}%
\bibitem [{\citenamefont {Bagrets}\ \emph {et~al.}(2016)\citenamefont
  {Bagrets}, \citenamefont {Celentano}, \citenamefont {Augieri}, \citenamefont
  {Nast},\ and\ \citenamefont {Weiss}}]{Bagrets2016}%
  \BibitemOpen
  \bibfield  {author} {\bibinfo {author} {\bibfnamefont {N.}~\bibnamefont
  {Bagrets}}, \bibinfo {author} {\bibfnamefont {G.}~\bibnamefont {Celentano}},
  \bibinfo {author} {\bibfnamefont {A.}~\bibnamefont {Augieri}}, \bibinfo
  {author} {\bibfnamefont {R.}~\bibnamefont {Nast}}, \ and\ \bibinfo {author}
  {\bibfnamefont {K.-P.}\ \bibnamefont {Weiss}},\ }\href
  {https://ieeexplore.ieee.org/abstract/document/7414426?casa_token=pWg08DlpDAAAAAAA:5fqhjCvPL1TQx_QIcOOGW6OHCffh1Lf_vE-ISoTzZcnl0s4_tfy5wI5nialylTHFRi8se_BOFV12}
  {\bibfield  {journal} {\bibinfo  {journal} {IEEE Transactions on Applied
  Superconductivity}\ }\textbf {\bibinfo {volume} {26}},\ \bibinfo {pages} {1}
  (\bibinfo {year} {2016})}\BibitemShut {NoStop}%
\bibitem [{\citenamefont {Fang}\ \emph {et~al.}(1996)\citenamefont {Fang},
  \citenamefont {Danyluk}, \citenamefont {Cha},\ and\ \citenamefont
  {Lanagan}}]{Fang1996}%
  \BibitemOpen
  \bibfield  {author} {\bibinfo {author} {\bibfnamefont {Y.}~\bibnamefont
  {Fang}}, \bibinfo {author} {\bibfnamefont {S.}~\bibnamefont {Danyluk}},
  \bibinfo {author} {\bibfnamefont {Y.}~\bibnamefont {Cha}}, \ and\ \bibinfo
  {author} {\bibfnamefont {M.~T.}\ \bibnamefont {Lanagan}},\ }\href
  {https://aip.scitation.org/doi/abs/10.1063/1.362694} {\bibfield  {journal}
  {\bibinfo  {journal} {Journal of applied physics}\ }\textbf {\bibinfo
  {volume} {79}},\ \bibinfo {pages} {947} (\bibinfo {year} {1996})}\BibitemShut
  {NoStop}%
\bibitem [{\citenamefont {Duckworth}(2001)}]{Duckworth2001}%
  \BibitemOpen
  \bibfield  {author} {\bibinfo {author} {\bibfnamefont {R.~C.}\ \bibnamefont
  {Duckworth}},\ }\href
  {https://fs.magnet.fsu.edu/~lee/asc/pdf_papers/theses/rcd01phd.pdf} {\enquote
  {\bibinfo {title} {Contact resistance and normal zone formation in coated
  yttrium barium copper oxide superconductors},}\ } (\bibinfo {year}
  {2001})\BibitemShut {NoStop}%
\bibitem [{\citenamefont {Usoskin}\ \emph {et~al.}(2002)\citenamefont
  {Usoskin}, \citenamefont {Issaev}, \citenamefont {Freyhardt}, \citenamefont
  {Leghissa}, \citenamefont {Oomen},\ and\ \citenamefont
  {Neumueller}}]{Usoskin2002}%
  \BibitemOpen
  \bibfield  {author} {\bibinfo {author} {\bibfnamefont {A.}~\bibnamefont
  {Usoskin}}, \bibinfo {author} {\bibfnamefont {A.}~\bibnamefont {Issaev}},
  \bibinfo {author} {\bibfnamefont {H.}~\bibnamefont {Freyhardt}}, \bibinfo
  {author} {\bibfnamefont {M.}~\bibnamefont {Leghissa}}, \bibinfo {author}
  {\bibfnamefont {M.}~\bibnamefont {Oomen}}, \ and\ \bibinfo {author}
  {\bibfnamefont {H.-W.}\ \bibnamefont {Neumueller}},\ }\href
  {https://www.sciencedirect.com/science/article/pii/S0921453402008729?casa_token=ARyPdrg5DvoAAAAA:ElSNsyZ-jrYTVBhfs-WJ5ZpWtbE5ToecV218PaqtFr8dR13JD0U3cgSyeluL1OmezpCyXgCIrpPn}
  {\bibfield  {journal} {\bibinfo  {journal} {Physica C: Superconductivity}\
  }\textbf {\bibinfo {volume} {372}},\ \bibinfo {pages} {857} (\bibinfo {year}
  {2002})}\BibitemShut {NoStop}%
\bibitem [{\citenamefont {Yu}\ \emph {et~al.}(1982)\citenamefont {Yu},
  \citenamefont {Eyssa},\ and\ \citenamefont {Zolliker}}]{yu1982}%
  \BibitemOpen
  \bibfield  {author} {\bibinfo {author} {\bibfnamefont {D.}~\bibnamefont
  {Yu}}, \bibinfo {author} {\bibfnamefont {Y.}~\bibnamefont {Eyssa}}, \ and\
  \bibinfo {author} {\bibfnamefont {P.}~\bibnamefont {Zolliker}},\ }in\ \href
  {https://link.springer.com/chapter/10.1007/978-1-4613-3542-9_70} {\emph
  {\bibinfo {booktitle} {Advances in Cryogenic Engineering Materials}}}\
  (\bibinfo  {publisher} {Springer},\ \bibinfo {year} {1982})\ pp.\ \bibinfo
  {pages} {701--709}\BibitemShut {NoStop}%
\bibitem [{\citenamefont {Wilson}(1983)}]{Wilson1983}%
  \BibitemOpen
  \bibfield  {author} {\bibinfo {author} {\bibfnamefont {M.}~\bibnamefont
  {Wilson}},\ }\href@noop {} {\emph {\bibinfo {title} {{Superconducting
  Magnets}}}}\ (\bibinfo {address} {Oxford},\ \bibinfo {year}
  {1983})\BibitemShut {NoStop}%
\bibitem [{\citenamefont {Lacroix}\ and\ \citenamefont
  {Sirois}(2014)}]{Lacroix2014b}%
  \BibitemOpen
  \bibfield  {author} {\bibinfo {author} {\bibfnamefont {C.}~\bibnamefont
  {Lacroix}}\ and\ \bibinfo {author} {\bibfnamefont {F.}~\bibnamefont
  {Sirois}},\ }\href
  {https://iopscience.iop.org/article/10.1088/0953-2048/27/3/035003/meta}
  {\bibfield  {journal} {\bibinfo  {journal} {Superconductor Science and
  Technology}\ }\textbf {\bibinfo {volume} {27}},\ \bibinfo {pages} {035003}
  (\bibinfo {year} {2014})}\BibitemShut {NoStop}%
\bibitem [{\citenamefont {Lacroix}\ \emph {et~al.}(2017)\citenamefont
  {Lacroix}, \citenamefont {Sirois},\ and\ \citenamefont
  {Lupien}}]{Lacroix2017}%
  \BibitemOpen
  \bibfield  {author} {\bibinfo {author} {\bibfnamefont {C.}~\bibnamefont
  {Lacroix}}, \bibinfo {author} {\bibfnamefont {F.}~\bibnamefont {Sirois}}, \
  and\ \bibinfo {author} {\bibfnamefont {J.~F.}\ \bibnamefont {Lupien}},\
  }\href {https://iopscience.iop.org/article/10.1088/1361-6668/aa684f/meta}
  {\bibfield  {journal} {\bibinfo  {journal} {Superconductor Science and
  Technology}\ }\textbf {\bibinfo {volume} {30}},\ \bibinfo {pages} {064004}
  (\bibinfo {year} {2017})}\BibitemShut {NoStop}%
\bibitem [{THE()}]{THEVA}%
  \BibitemOpen
  \href@noop {} {\enquote {\bibinfo {title} {{THEVA GmbH} website},}\ }\bibinfo
  {howpublished} {\url{https://www.theva.com/}},\ \bibinfo {note} {accessed:
  2020-09-06}\BibitemShut {NoStop}%
\bibitem [{\citenamefont {Bauer}\ \emph
  {et~al.}(1999{\natexlab{a}})\citenamefont {Bauer}, \citenamefont {Semerad},\
  and\ \citenamefont {Kinder}}]{Bauer1999}%
  \BibitemOpen
  \bibfield  {author} {\bibinfo {author} {\bibfnamefont {M.}~\bibnamefont
  {Bauer}}, \bibinfo {author} {\bibfnamefont {R.}~\bibnamefont {Semerad}}, \
  and\ \bibinfo {author} {\bibfnamefont {H.}~\bibnamefont {Kinder}},\ }\href
  {https://ieeexplore.ieee.org/abstract/document/784678} {\bibfield  {journal}
  {\bibinfo  {journal} {IEEE transactions on applied superconductivity}\
  }\textbf {\bibinfo {volume} {9}},\ \bibinfo {pages} {1502} (\bibinfo {year}
  {1999}{\natexlab{a}})}\BibitemShut {NoStop}%
\bibitem [{\citenamefont {Bauer}\ \emph
  {et~al.}(1999{\natexlab{b}})\citenamefont {Bauer}, \citenamefont {Metzger},
  \citenamefont {Semerad}, \citenamefont {Berberich},\ and\ \citenamefont
  {Kinder}}]{Bauer2000}%
  \BibitemOpen
  \bibfield  {author} {\bibinfo {author} {\bibfnamefont {M.}~\bibnamefont
  {Bauer}}, \bibinfo {author} {\bibfnamefont {R.}~\bibnamefont {Metzger}},
  \bibinfo {author} {\bibfnamefont {R.}~\bibnamefont {Semerad}}, \bibinfo
  {author} {\bibfnamefont {P.}~\bibnamefont {Berberich}}, \ and\ \bibinfo
  {author} {\bibfnamefont {H.}~\bibnamefont {Kinder}},\ }\href
  {https://www.cambridge.org/core/journals/mrs-online-proceedings-library-archive/article/inclined-substrate-deposition-by-evaporation-of-magnesium-oxide-for-coated-conductors/57946BEF6E572357052A187FA5DFFBD2}
  {\bibfield  {journal} {\bibinfo  {journal} {MRS Online Proceedings Library
  Archive}\ }\textbf {\bibinfo {volume} {585}} (\bibinfo {year}
  {1999}{\natexlab{b}})}\BibitemShut {NoStop}%
\bibitem [{\citenamefont {Metzger}\ \emph {et~al.}(2001)\citenamefont
  {Metzger}, \citenamefont {Bauer}, \citenamefont {Numssen}, \citenamefont
  {Semerad}, \citenamefont {Berberich},\ and\ \citenamefont
  {Kinder}}]{Metzger2001}%
  \BibitemOpen
  \bibfield  {author} {\bibinfo {author} {\bibfnamefont {R.}~\bibnamefont
  {Metzger}}, \bibinfo {author} {\bibfnamefont {M.}~\bibnamefont {Bauer}},
  \bibinfo {author} {\bibfnamefont {K.}~\bibnamefont {Numssen}}, \bibinfo
  {author} {\bibfnamefont {R.}~\bibnamefont {Semerad}}, \bibinfo {author}
  {\bibfnamefont {P.}~\bibnamefont {Berberich}}, \ and\ \bibinfo {author}
  {\bibfnamefont {H.}~\bibnamefont {Kinder}},\ }\href
  {https://ieeexplore.ieee.org/document/919651} {\bibfield  {journal} {\bibinfo
   {journal} {IEEE transactions on applied superconductivity}\ }\textbf
  {\bibinfo {volume} {11}},\ \bibinfo {pages} {2826} (\bibinfo {year}
  {2001})}\BibitemShut {NoStop}%
\bibitem [{\citenamefont {Zilbauer}\ \emph {et~al.}(2006)\citenamefont
  {Zilbauer}, \citenamefont {Berberich}, \citenamefont {L{\"u}mkemann},
  \citenamefont {Numssen}, \citenamefont {Wassner},\ and\ \citenamefont
  {Sigl}}]{Zilbauer2006}%
  \BibitemOpen
  \bibfield  {author} {\bibinfo {author} {\bibfnamefont {T.}~\bibnamefont
  {Zilbauer}}, \bibinfo {author} {\bibfnamefont {P.}~\bibnamefont {Berberich}},
  \bibinfo {author} {\bibfnamefont {A.}~\bibnamefont {L{\"u}mkemann}}, \bibinfo
  {author} {\bibfnamefont {K.}~\bibnamefont {Numssen}}, \bibinfo {author}
  {\bibfnamefont {T.}~\bibnamefont {Wassner}}, \ and\ \bibinfo {author}
  {\bibfnamefont {G.}~\bibnamefont {Sigl}},\ }\href
  {https://iopscience.iop.org/article/10.1088/0953-2048/19/11/005/meta}
  {\bibfield  {journal} {\bibinfo  {journal} {Superconductor Science and
  Technology}\ }\textbf {\bibinfo {volume} {19}},\ \bibinfo {pages} {1118}
  (\bibinfo {year} {2006})}\BibitemShut {NoStop}%
\bibitem [{\citenamefont {Kinder}\ \emph {et~al.}(1997)\citenamefont {Kinder},
  \citenamefont {Berberich}, \citenamefont {Prusseit}, \citenamefont
  {Rieder-Zecha}, \citenamefont {Semerad},\ and\ \citenamefont
  {Utz}}]{Kinder1997}%
  \BibitemOpen
  \bibfield  {author} {\bibinfo {author} {\bibfnamefont {H.}~\bibnamefont
  {Kinder}}, \bibinfo {author} {\bibfnamefont {P.}~\bibnamefont {Berberich}},
  \bibinfo {author} {\bibfnamefont {W.}~\bibnamefont {Prusseit}}, \bibinfo
  {author} {\bibfnamefont {S.}~\bibnamefont {Rieder-Zecha}}, \bibinfo {author}
  {\bibfnamefont {R.}~\bibnamefont {Semerad}}, \ and\ \bibinfo {author}
  {\bibfnamefont {B.}~\bibnamefont {Utz}},\ }\href
  {https://www.sciencedirect.com/science/article/pii/S092145349700230X}
  {\bibfield  {journal} {\bibinfo  {journal} {Physica C: Superconductivity}\
  }\textbf {\bibinfo {volume} {282}},\ \bibinfo {pages} {107} (\bibinfo {year}
  {1997})}\BibitemShut {NoStop}%
\bibitem [{\citenamefont {Russek}\ \emph {et~al.}(1994)\citenamefont {Russek},
  \citenamefont {Sanders}, \citenamefont {Roshko},\ and\ \citenamefont
  {Ekin}}]{Russek1994}%
  \BibitemOpen
  \bibfield  {author} {\bibinfo {author} {\bibfnamefont {S.~E.}\ \bibnamefont
  {Russek}}, \bibinfo {author} {\bibfnamefont {S.~C.}\ \bibnamefont {Sanders}},
  \bibinfo {author} {\bibfnamefont {A.}~\bibnamefont {Roshko}}, \ and\ \bibinfo
  {author} {\bibfnamefont {J.}~\bibnamefont {Ekin}},\ }\href
  {http://scitation.aip.org/content/aip/journal/apl/64/26/10.1063/1.111192}
  {\bibfield  {journal} {\bibinfo  {journal} {Applied physics letters}\
  }\textbf {\bibinfo {volume} {64}},\ \bibinfo {pages} {3649} (\bibinfo {year}
  {1994})}\BibitemShut {NoStop}%
\bibitem [{\citenamefont {Ekin}(2006)}]{Ekin2006}%
  \BibitemOpen
  \bibfield  {author} {\bibinfo {author} {\bibfnamefont {J.}~\bibnamefont
  {Ekin}},\ }\href@noop {} {\emph {\bibinfo {title} {{Experimental techniques
  for low-temperature \\ measurements: cryostat design, material properties and
  superconductor critical-current testing}}}}\ (\bibinfo  {publisher} {Oxford
  university press},\ \bibinfo {year} {2006})\BibitemShut {NoStop}%
\bibitem [{\citenamefont {Fournier-Lupien}\ \emph {et~al.}(2020)\citenamefont
  {Fournier-Lupien}, \citenamefont {Del~Vecchio}, \citenamefont {Lacroix},\
  and\ \citenamefont {Sirois}}]{Fournier-Lupien2020a}%
  \BibitemOpen
  \bibfield  {author} {\bibinfo {author} {\bibfnamefont {J.-H.}\ \bibnamefont
  {Fournier-Lupien}}, \bibinfo {author} {\bibfnamefont {P.}~\bibnamefont
  {Del~Vecchio}}, \bibinfo {author} {\bibfnamefont {C.}~\bibnamefont
  {Lacroix}}, \ and\ \bibinfo {author} {\bibfnamefont {F.}~\bibnamefont
  {Sirois}},\ }\href
  {https://iopscience.iop.org/article/10.1088/1361-6668/aba543/meta} {\bibfield
   {journal} {\bibinfo  {journal} {Superconductor Science and Technology}\
  }\textbf {\bibinfo {volume} {33}},\ \bibinfo {pages} {115014} (\bibinfo
  {year} {2020})}\BibitemShut {NoStop}%
\bibitem [{\citenamefont {Lacroix}\ \emph {et~al.}(2013)\citenamefont
  {Lacroix}, \citenamefont {Fournier-Lupien}, \citenamefont {McMeekin},\ and\
  \citenamefont {Sirois}}]{Lacroix2013a}%
  \BibitemOpen
  \bibfield  {author} {\bibinfo {author} {\bibfnamefont {C.}~\bibnamefont
  {Lacroix}}, \bibinfo {author} {\bibfnamefont {J.-H.}\ \bibnamefont
  {Fournier-Lupien}}, \bibinfo {author} {\bibfnamefont {K.}~\bibnamefont
  {McMeekin}}, \ and\ \bibinfo {author} {\bibfnamefont {F.}~\bibnamefont
  {Sirois}},\ }\href {https://ieeexplore.ieee.org/abstract/document/6409995}
  {\bibfield  {journal} {\bibinfo  {journal} {IEEE transactions on applied
  superconductivity}\ }\textbf {\bibinfo {volume} {23}},\ \bibinfo {pages}
  {4701605} (\bibinfo {year} {2013})}\BibitemShut {NoStop}%
\bibitem [{Note1()}]{Note1}%
  \BibitemOpen
  \bibinfo {note} {Hypertac MLPI-30-4-L-SMT}\BibitemShut {NoStop}%
\bibitem [{\citenamefont {{Keithley Instruments}}(2004)}]{Keithley2004}%
  \BibitemOpen
  \bibfield  {author} {\bibinfo {author} {\bibfnamefont {I.}~\bibnamefont
  {{Keithley Instruments}}},\ }\href
  {https://web.mit.edu/8.13/8.13d/manuals/LowLevMsHandbk.pdf} {\emph {\bibinfo
  {title} {{Low Level Measurements Handbook, 6th Edition, Precision DC Current,
  Voltage and Resistance Measurements}}}},\ \bibinfo {type} {Tech. Rep.}\
  (\bibinfo {year} {2004})\BibitemShut {NoStop}%
\bibitem [{\citenamefont {Fournier-Lupien}\ \emph
  {et~al.}(2018{\natexlab{a}})\citenamefont {Fournier-Lupien}, \citenamefont
  {Lacroix}, \citenamefont {Hellmann}, \citenamefont {Huh}, \citenamefont
  {Pfeiffer},\ and\ \citenamefont {Sirois}}]{Fournier-Lupien2018a}%
  \BibitemOpen
  \bibfield  {author} {\bibinfo {author} {\bibfnamefont {J.}~\bibnamefont
  {Fournier-Lupien}}, \bibinfo {author} {\bibfnamefont {C.}~\bibnamefont
  {Lacroix}}, \bibinfo {author} {\bibfnamefont {S.}~\bibnamefont {Hellmann}},
  \bibinfo {author} {\bibfnamefont {J.}~\bibnamefont {Huh}}, \bibinfo {author}
  {\bibfnamefont {K.}~\bibnamefont {Pfeiffer}}, \ and\ \bibinfo {author}
  {\bibfnamefont {F.}~\bibnamefont {Sirois}},\ }\href
  {https://iopscience.iop.org/article/10.1088/1361-6668/aae2cd/meta} {\bibfield
   {journal} {\bibinfo  {journal} {Superconductor Science and Technology}\
  }\textbf {\bibinfo {volume} {31}},\ \bibinfo {pages} {125019} (\bibinfo
  {year} {2018}{\natexlab{a}})}\BibitemShut {NoStop}%
\bibitem [{\citenamefont {Brandt}(1996)}]{Brandt1996}%
  \BibitemOpen
  \bibfield  {author} {\bibinfo {author} {\bibfnamefont {E.~H.}\ \bibnamefont
  {Brandt}},\ }\href
  {https://journals.aps.org/prb/abstract/10.1103/PhysRevB.54.4246} {\bibfield
  {journal} {\bibinfo  {journal} {Physical review B}\ }\textbf {\bibinfo
  {volume} {54}},\ \bibinfo {pages} {4246} (\bibinfo {year}
  {1996})}\BibitemShut {NoStop}%
\bibitem [{\citenamefont {Fournier-Lupien}\ \emph
  {et~al.}(2018{\natexlab{b}})\citenamefont {Fournier-Lupien}, \citenamefont
  {Lacroix}, \citenamefont {Hellmann}, \citenamefont {Huh}, \citenamefont
  {Pfeiffer},\ and\ \citenamefont {Sirois}}]{fournier2018}%
  \BibitemOpen
  \bibfield  {author} {\bibinfo {author} {\bibfnamefont {J.}~\bibnamefont
  {Fournier-Lupien}}, \bibinfo {author} {\bibfnamefont {C.}~\bibnamefont
  {Lacroix}}, \bibinfo {author} {\bibfnamefont {S.}~\bibnamefont {Hellmann}},
  \bibinfo {author} {\bibfnamefont {J.}~\bibnamefont {Huh}}, \bibinfo {author}
  {\bibfnamefont {K.}~\bibnamefont {Pfeiffer}}, \ and\ \bibinfo {author}
  {\bibfnamefont {F.}~\bibnamefont {Sirois}},\ }\href
  {https://iopscience.iop.org/article/10.1088/1361-6668/aae2cd} {\bibfield
  {journal} {\bibinfo  {journal} {Superconductor Science and Technology}\
  }\textbf {\bibinfo {volume} {31}},\ \bibinfo {pages} {125019} (\bibinfo
  {year} {2018}{\natexlab{b}})}\BibitemShut {NoStop}%
\end{thebibliography}%

\end{document}